\documentclass[11pt]{article}
\usepackage{jheppub}
\usepackage{graphicx}
\usepackage{amsmath}
\usepackage{datetime}

\numberwithin{equation}{section}

\newcommand{\be}{\begin{equation}}
\newcommand{\ee}{\end{equation}}
\newcommand{\bea}{\begin{eqnarray}}
\newcommand{\eea}{\end{eqnarray}}
\newcommand{\bear}{\begin{eqnarray}}
\newcommand{\eear}{\end{eqnarray}}
\newcommand{\beas}{\begin{eqnarray*}}
\newcommand{\p}{\partial}
\newcommand{\eeas}{\end{eqnarray*}}
\newcommand{\ba}{\begin{array}}
\newcommand{\ea}{\end{array}}

\newcommand{\nn}{\nonumber}
\newcommand{\del}{\nabla}

\def\half{\frac{1}{2}}
\def\D{\Delta}

\def\r{\rightarrow}
\def\d{\delta}
\def\l{\lambda}
\def\B{B}
\newcommand{\bi}{\begin{itemize}}
\newcommand{\ei}{\end{itemize}}

\newcommand{\Dh}[1]{\delta \langle H_{#1} \rangle}

\renewcommand{\a}{\alpha}
\renewcommand{\b}{\beta}

\newcommand{\s}{\sigma}

\newcommand{\tr}{\operatorname{tr}}
\newcommand{\pd}[2][1]{\ifnum#1=1 \frac{\partial}{\partial {#2}} \else
  \frac{\partial^#1}{\partial {#2}^{#1}}\fi}
\newcommand{\dpd}[2][1]{\ifnum#1=1 \dfrac{\partial}{\partial {#2}} \else
  \frac{\partial^#1}{\partial {#2}^{#1}}\fi}
\newcommand{\td}[2][1]{\ifnum#1=1 \frac{d}{d{#2}} \else
  \frac{d^#1}{d{#2}^{#1}}\fi}

\newcommand{\g}{\gamma}

\newcommand{\e}{\varepsilon}

\renewcommand{\(}{\left(}
\renewcommand{\)}{\right)}

\newcommand{\nbox}{{\,\lower0.9pt\vbox{\hrule \hbox{\vrule height 0.2 cm \hskip 0.19 cm \vrule height 0.2 cm}\hrule}\,}}

\def\href#1#2{#2}

\newcommand{\mt}[1]{\textrm{\tiny #1}}

\newcommand{\Gn}{G_\mt{N}}
\newcommand{\reef}[1]{(\ref{#1})}

\newcommand{\ie}{{\it i.e.,}\ }

\newcommand{\labell}[1]{\qquad\mt{#1}\qquad\label{#1}}
\renewcommand{\labell}[1]{\label{#1}}

\newcommand{\bL}{\mathbf{L}}
\newcommand{\bTheta}{\mathbf{\Theta}}
\newcommand{\bE}{\mathbf{E}}

\newcommand{\bJ}{\mathbf{J}}
\newcommand{\bQ}{\mathbf{Q}}
\newcommand{\bX}{\mathbf{X}}
\newcommand{\bepsilon}{\boldsymbol{\varepsilon}}
\newcommand{\bomega}{\pmb{\omega}}
\newcommand{\bC}{\mathbf{C}}
\newcommand{\bV}{\boldsymbol{\varepsilon}}
\newcommand{\bchi}{\boldsymbol{\chi}}

\newcommand{\bW}{\mathbf{W}}
\newcommand{\bY}{\mathbf{Y}}
\newcommand{\modu}{modular\ }
\newcommand{\Sw}{S_\mt{Wald}}
\newcommand{\tl}{{\tilde\ell}}

\title{Gravitation from entanglement in holographic CFTs}

\author[a]{Thomas Faulkner,}
\author[b]{Monica Guica,}
\author[c]{Thomas Hartman,}
\author[d]{Robert C. Myers}
\author[e]{\\and Mark Van Raamsdonk}

\affiliation[a]{Institute for Advanced Study, Princeton, NJ, 08540 \vspace{2mm}}

\affiliation[b]{Department of Physics and Astronomy, University of Pennsylvania\\ 209 S. 33rd St., Philadelphia, PA 19104-6396, USA \vspace{2mm}}

\affiliation[c]{Kavli Institute for Theoretical Physics, University of California\\
Santa Barbara, CA 93106-4030 USA \vspace{2mm}}

\affiliation[d]{Perimeter Institute for Theoretical Physics\\
31 Caroline Street N., Waterloo, Ontario N2L 2Y5, Canada\vspace{2mm}}

\affiliation[e]{Department of Physics and Astronomy,
University of British Columbia\\
6224 Agricultural Road, Vancouver, B.C. V6T 1W9, Canada \vspace{4mm}}

\emailAdd{faulkner@ias.edu}
\emailAdd{mguica@sas.upenn.edu}
\emailAdd{thartman@kitp.ucsb.edu}
\emailAdd{rmyers@perimeterinstitute.ca}
\emailAdd{mav@phas.ubc.ca}

\preprint{\today, \currenttime}

\abstract{

\vspace{0.5 cm}

\noindent
Entanglement entropy obeys a `first law', an exact quantum generalization of the ordinary first law of
thermodynamics.  In any CFT with a semiclassical holographic dual, this first law has an interpretation in the dual gravitational theory as a constraint on the spacetimes dual to CFT states. For small perturbations around the CFT vacuum state, we show that the set of such constraints for all ball-shaped spatial regions in the CFT is exactly equivalent to the requirement that the dual geometry satisfy the gravitational equations of motion, linearized about pure AdS. For theories with entanglement entropy computed by the Ryu-Takayanagi formula $S = {\cal A} /(4\Gn)$, we obtain the linearized Einstein equations. For theories in which the vacuum entanglement entropy for a ball is computed by more general Wald functionals, we obtain the linearized equations for the associated higher-curvature theories. Using the first law, we also derive the holographic dictionary for the stress tensor, given the holographic formula for entanglement entropy. This method provides a simple alternative to holographic renormalization for computing the stress tensor expectation value in arbitrary higher derivative gravitational theories.}

\begin{document}

\maketitle

\section{Introduction}

According to the AdS/CFT correspondence, spacetime and gravitational physics in AdS emerge from the dynamics of certain strongly-coupled conformal field theories with a large number of degrees of freedom. A central question is to understand why and how this happens.
In recent work, it has been suggested that the physics of quantum entanglement plays an essential role, e.g. \cite{rt1,rt2,mav1,swing,arch}. This was motivated in part by the importance of quantum entanglement for understanding quantum phases of matter in condensed matter systems \cite{cmt}. Ryu and Takayanagi have proposed \cite{rt1,rt2} that entanglement entropy, one measure of entanglement between subsets of degrees of freedom in general quantum systems, provides a direct window into the emergent spacetime geometry, giving the areas of certain extremal surfaces. This provides a quantitative connection between CFT entanglement and the dual spacetime geometry. Recently, this connection has been utilized to understand the emergence of spacetime dynamics (i.e. gravity) from the CFT physics \cite{eom}. Making use of a `first law' for entanglement entropy derived in \cite{relative}, it was shown \cite{eom} that in any holographic theory for which the Ryu-Takayanagi prescription computes the entanglement entropy of the boundary CFT, spacetimes dual to small perturbations of the CFT vacuum state must satisfy Einstein's equations linearized around pure AdS spacetime.

In this paper, we provide further insight into the results of \cite{relative,eom} and extend them to general holographic CFTs, for which the classical bulk equations may include terms at higher order in the curvatures or derivatives. We show further that the first law for entanglement entropy in the CFT can be understood as the microscopic origin of a particular case of the first law of black hole thermodynamics, applied to AdS-Rindler horizons. We begin with a brief review of some essential background before summarizing our main results.

\subsubsection*{The `first law' of entanglement entropy}

The crucial piece of CFT physics giving rise to linearized gravitational equations in the dual theory is a `first law' of entanglement entropy,
\be\labell{firstlaw0}
\delta S_A = \delta \langle H_A\rangle \;
\ee
equating the first order variation in the entanglement entropy for a spatial region $A$ with the first order variation in the expectation value of $H_A$, the modular (or entanglement) Hamiltonian. The latter operator is
defined as the logarithm of the unperturbed state, i.e.
$\rho_A\simeq e^{-H_A}$ --- see section \ref{ss:firstlaw} for further details. The first law was derived in \cite{relative}\footnote{Related observations had been made independently using
various holographic calculations, e.g. \cite{therm}.} as a special case of a more general result for finite perturbations
\be\labell{inequality}
\Delta S_A \le \Delta \langle H_A\rangle \;
\ee
obtained using the positivity of `relative entropy'.\footnote{Relative entropy can be viewed as a statistical measure 
of the distance between two states
(i.e. density matrices) in the same Hilbert space --- e.g. see \cite{wehrl,vedral1} for reviews.} A more direct demonstration of \reef{firstlaw0} is reviewed in section \ref{ss:firstlaw} below.

In general, the modular Hamiltonian $H_A$ is a complicated object that cannot be expressed as an integral of
local operators. However, starting from the vacuum state of a CFT in flat space and taking $A$ to be a ball-shaped spatial region of radius $R$ centered at $x_0$, denoted $B(R,x_0)$, the modular Hamiltonian is given by a simple integral \cite{chm}
 \be\labell{modhcft0}
H_B = 2\pi\ \int_{B(R,x_0)}\!\!\!\!\!\!\!\!  d^{d-1}x\  \frac{R^2 -
|\vec{x} - \vec{x}_0|^2}{2R}\ T_{tt} \ ,
 \ee
of the energy density over the interior of the sphere (weighted by a certain spatial profile). Thus, given any perturbation to the CFT vacuum  we have for any ball-shaped region
\be\labell{firstlaw1}
\delta S_B = 2\pi\ \int_{B(R,x_0)}\!\!\!\!\!\!\!\!  d^{d-1}x\  \frac{R^2 -
|\vec{x} - \vec{x}_0|^2}{2R}\ \d \langle  T_{tt}\rangle \ ,  \;
\ee
where $H_B$ and $S_B$ denote the  modular Hamiltonian and the entanglement entropy for a ball, respectively.

\medskip

\subsubsection*{The holographic interpretation}

For conformal field theories with a gravity dual, the first law for ball-shaped regions can be translated into a geometrical constraint obeyed by any spacetime dual to a small perturbation of the CFT vacuum. To understand this, we first recall the holographic interpretation of entanglement entropy and energy density in the general case (see section \ref{ss:holographic} for more details).

As shown by \cite{chm} in deriving \reef{modhcft0}, the vacuum entanglement entropy of a CFT for a ball-shaped region in flat space can be reinterpreted as the thermal entropy of the CFT on a hyperbolic cylinder at temperature set by the hyperbolic space curvature scale, by relating the two backgrounds with a conformal mapping. For a holographic CFT, the latter thermal entropy may then be calculated as the horizon entropy
of the ``black hole''  dual to this thermal state on hyperbolic space. In this case, the black hole is simply a Rindler wedge (which we call the AdS-Rindler patch) of the original pure AdS space, as shown in figure \ref{hyperbolic2}.
If the gravitational theory in the bulk
is Einstein gravity, then the horizon entropy is given by the usual
Bekenstein-Hawking formula, $S_\mt{BH}={\cal A}/(4\Gn)$, and this construction
\cite{chm} provides a derivation of the Ryu-Takayanagi prescription \cite{rt1}
for a spherical entangling surface.\footnote{Recently, this approach was
extended to a general argument for the Ryu-Takayanagi prescription for arbitrary
entangling surfaces in time-independent (and some special time-dependent) backgrounds \cite{aitor}.} However, we note that the same analysis
applies for any classical and covariant gravity theory in the bulk, in
which case the horizon entropy is given by Wald's formula
\cite{wald0,ted1,iyer1}
 \be\labell{waldent0}
\Sw = - 2\pi \int_{\mathcal{H}} d^{n}\sigma\ \sqrt{h}\
\frac{\delta {\cal L}}{\delta R^{ab}{}_{cd}}\, n^{ab}\,n_{cd} \ ,
 \ee
where $\cal L$ denotes the gravitational Lagrangian and $n_{ab}$ is the
binormal to the horizon $\mathcal{H}$.

To summarize, in general holographic theories, entanglement entropy in the vacuum state for a ball-shaped region $B$ is computed by the Wald functional applied to
the horizon of the AdS-Rindler patch associated with $B$.
We will argue in section \ref{ss:holographic} that this should remain true for perturbations to the vacuum state,
so the left side of \reef{firstlaw1} computes the change in entropy of the
AdS-Rindler horizon
under a small variation of the CFT state.
Meanwhile, the expectation value of the stress tensor is related to the asymptotic behaviour of the metric,
so the right side of \reef{firstlaw1} may be expressed as an integral involving the asymptotic metric over a ball-shaped region of the boundary. In section \ref{ss:holographic}, we show that this integral may be interpreted as the variation in energy of
the AdS-Rindler spacetime.
Thus, the gravity version of the entanglement first law \reef{firstlaw1} may be interpreted as a first law
for AdS-Rindler spacetimes.
At a technical level, this represents a non-local constraint on the spacetime fields, equating an integral involving the asymptotic metric perturbation over a boundary surface to an integral involving the bulk metric perturbation (and possibly matter fields) over a bulk surface.

\subsubsection*{Main results}

Our first main result, presented in section \ref{s:forward}, is that this first law for
AdS-Rindler spacetimes,
i.e. the gravitational version of \reef{firstlaw1}, is a special case of a first law proved by Iyer and Wald for stationary
spacetimes
with bifurcate Killing horizons (i.e. at finite temperature) in general classical theories of gravity. According to Iyer and Wald, for any perturbation of a stationary
 background  that satisfies the linearized equations of motion following from some Lagrangian, the first law holds provided we define horizon
 entropy using the Wald functional \reef{waldent0} associated with this Lagrangian. Thus, the CFT result \reef{firstlaw1} can be seen as an exact quantum version of the Iyer-Wald first law, at least for the case of AdS-Rindler horizons.

Our second result, presented in sections \ref{s:converse} and \ref{s:detail}, provides a converse to the theorem of Iyer and Wald. In AdS space, we can associate
an AdS-Rindler patch
to any ball-shaped spatial region on the boundary in any Lorentz frame, as in figure \ref{hyperbolic2}. An arbitrary perturbation to the AdS metric can be understood as a perturbation to each of these
Rindler patches.
We show that if the first law is satisfied for
every AdS-Rindler patch,
then the perturbation must satisfy the linearized gravitational equations. Thus, the set of non-local constraints (one for each ball-shaped region in each Lorentz frame) implied by \reef{firstlaw1} {\it is equivalent} to the set of local gravitational equations.

The result in the previous paragraph --
that the first law for AdS-Rindler patches implies the linearized gravitational equations --
is completely independent of AdS/CFT and holds for any classical theory of gravity in AdS. However, since for holographic CFTs this gravitational first law is implied by the entanglement first law, we conclude that the linearized gravitational equations for the dual spacetime can be derived from any holographic CFT, given the entanglement functional. This extends the results of \cite{eom} to general holographic CFTs.

As a further application of the entanglement first law, we point out (see section \ref{ss:derivet}) that eq.~\reef{firstlaw0}, 
applied to infinitesimal balls, can be used to deduce the `holographic stress tensor,' i.e. the gravitational quantity that computes the expectation value of the CFT stress tensor, given the holographic prescription for computing entanglement entropy. This provides a simple alternative approach to the usual holographic renormalization procedure, as we illustrate with examples in section \ref{hcurv}. Finally, we show that eq.~\reef{firstlaw0} also provides information about the operators in the boundary theory corresponding to additional degrees of freedom that can be associated with the metric in the context of higher derivative gravity.

We conclude in section \ref{discuss} with a brief discussion of our results. In particular, we discuss the relation of our work to the work of Jacobson \cite{revelation}, who obtained gravitational equations by considering a gravitational first law applied to local Rindler horizons.

\section{Background} \labell{back}

In this section, we review some basic facts about entanglement
entropy, \modu Hamiltonians and their holographic interpretation. In section \ref{ss:firstlaw}, following
\cite{relative}, we review the first law-like relation $\delta S_A = \Dh A$ satisfied by entanglement entropy, specializing to entanglement for ball-shaped regions in a conformal field theory in section \ref{ss:firstlawCFT}. In section
\ref{ss:holographic}, we review the bulk interpretation of $S_A$ and $\langle H_A\rangle$ in a
holographic CFT.

\subsection{The first law of entanglement entropy}
\labell{ss:firstlaw}

For any state in a general quantum system, the state of a subsystem $A$ is described by a reduced density matrix $\rho_A = \tr_{\bar{A}}\rho_{\rm total}$, where $\rho_{\rm total}$ is the density matrix describing the global state of the full system and $\bar{A}$ is the complement of $A$. The entanglement of this subsystem with the rest of the system may be quantified by the entanglement entropy $S_A$, defined as the von Neumann entropy
 \be\labell{EE}
S_A = -\tr \rho_A \log \rho_A \
 \ee
of the density matrix $\rho_A$.

Since the reduced density matrix $\rho_A$ is both hermitian and positive (semi)definite, it
can be expressed as
 \be \labell{modh}
\rho_A = \frac{e^{-H_A}}{\tr\!\left( e^{-H_A} \right)} \ ,
 \ee
where the Hermitian operator $H_A$ is known as the \modu Hamiltonian. The denominator
is included on the right in the expression above to ensure that the reduced density matrix has unit trace. Note the eq.~\reef{modh} only defines $H_A$ up to an additive constant.

Now, consider any infinitesimal variation to the state of the system. The first order variation\footnote{Here and below, the variations are defined by considering a one-parameter family of states $|\Psi(\lambda) \rangle$ such that $|\Psi(0) \rangle = |0 \rangle$. The variation $\delta {\cal O}$ of any quantity associated with $|\Psi \rangle$ is then defined by $\delta {\cal O} = \partial_\lambda {\cal O}(\lambda)|_{\lambda = 0}$. \label{footy}} of the entanglement entropy \reef{EE} is
given by
 \bea
\d S_A &=& -\tr\!\( \d\rho_A \log \rho_A\)-\tr\!\( \rho_A\, \rho_A^{-1}\d\rho_A\)
\nn\\
\labell{derfirst}
&=& \tr\!\( \d\rho_A\, H_A\)-\tr\!\( \d\rho_A\) \ .
 \eea
Since the the trace of the reduced density matrix equals one by definition,
we must have $\tr\!\( \d\rho_A\)=0$. Hence, the variation of the entanglement entropy obeys
\be
\labell{firstlaw}
\d S_A =\d  \langle H_A \rangle\  ,
\ee
where $H_A$ is the modular Hamiltonian associated with the original unperturbed state.

In cases where we start with a thermal state $\rho_A = e^{-\beta H}/\tr(e^{-\beta H})$, equation (\ref{firstlaw}) gives $\d  \langle H \rangle\ = T \d S_A$, an exact quantum version of the first law of thermodynamics. Thus, \reef{firstlaw} represents a generalization of the first law of thermodynamics valid for arbitrary perturbations to arbitrary (non-equilibrium) states.

\subsection{The first law in conformal field theories}
\labell{ss:firstlawCFT}

\begin{figure}
\centering
\includegraphics[width=0.8\textwidth]{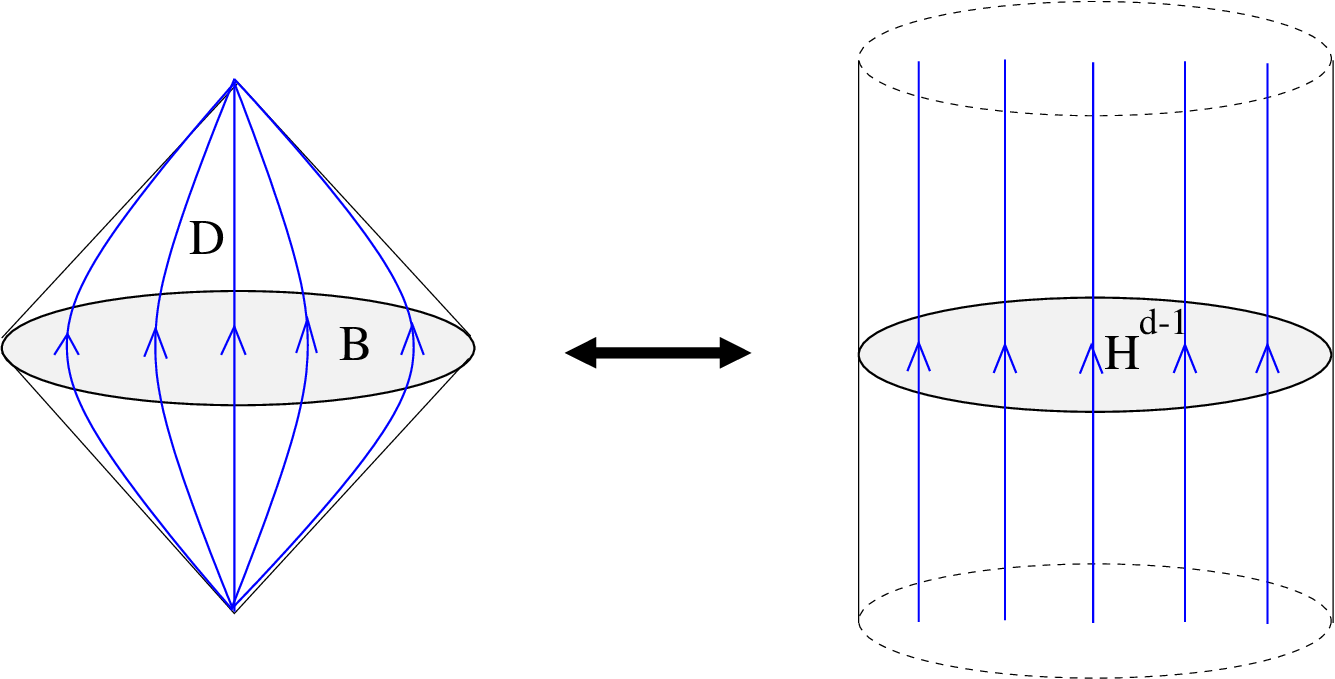}\\
\caption{Causal development $\mathcal{D}$ (left) of a ball-shaped region $B$ on a spatial slice of Minkowski space, showing the evolution generated by $H_B$. A conformal transformation maps $D$ to a hyperbolic cylinder $H^{d-1} \times$ time (right), taking $H_B$ to the ordinary Hamiltonian for the CFT on $H^{d-1}$.}
\label{conformal}
\end{figure}

We now specialize to the case of local quantum field theories. Here, for any fixed Cauchy surface, the field configurations on this time slice are representative of the Hilbert space of the underlying quantum theory. We can then define a subsystem A by introducing a smooth boundary or `entangling surface', which divides the Cauchy surface into two separate regions, $A$ and $\bar{A}$; the local fields in the region $A$ define a subsystem.

In general, the relation \reef{firstlaw} is of limited use. For a general quantum field theory, a general state, and a general region $A$,
the \modu Hamiltonian is not known and there is no known practical method to compute
it. Typically, $H_A$ is expected to be a complicated non-local operator. However,
there are a few situations where the \modu Hamiltonian has been established to have a simple form
as the integral of a local operator, and in which it generates a simple geometric flow.

One example is when we consider a conformal field theory in its vacuum state, $\rho_{\rm
total} = |0\rangle \langle 0|$ in $d$-dimensional Minkowski space, and
choose the region $A$ to be a ball $B(R,x_0)$ of radius $R$ on
a time slice $t=t_0$ and centered at $x^i=x^i_0$.\footnote{Our notation for the flat space
coordinates will be $x^\mu = (t,\vec{x})$ or $(t, x^i)$ where $i=1\dots d-1$, while $x^a = (z,t,\vec{x})$ denotes a coordinate on $AdS_{d+1}$.}
For this particular case, the \modu Hamiltonian takes the simple form \cite{Hislop:1981uh,chm}
 \be\labell{modhcft}
H_\B = 2\pi\ \int_{\B(R,x_0)}\!\!\!\!\!\!\!\! d^{d-1}x\  \frac{R^2 -
|\vec{x}-\vec{x}_0|^2}{2R}\ T_{tt}(t_0,\vec{x}) \ ,
 \ee
where $T_{\mu \nu}$ is the stress tensor.

To understand the origin of this expression, we recall that the causal development\footnote{The causal development $\mathcal{D}$ of the ball (also known as the domain of dependence) comprises all points $p$ for which all causal curves through $p$ necessarily intersect  $\B(R,\vec{x}_0)$.} $\mathcal{D}$ of $B$ is related by a conformal transformation to a hyperbolic cylinder ${\mathcal{H}} = H^{d-1} \times\, R_\tau$ (time) as shown in figure \ref{conformal}. As argued in \cite{chm}, this transformation induces a map of CFT states that takes the vacuum density matrix on $B$ to the thermal density matrix $\rho_{\mathcal{H}} \sim \exp(-2 \pi R H_\tau)$ for the CFT on hyperbolic space, where $R$ is the curvature radius of the hyperbolic space and $H_\tau$ is the CFT Hamiltonian generating time translations in ${\mathcal{H}}$.
The modular Hamiltonian for $\rho_{\mathcal{H}}$ is then just $2 \pi R H_\tau$. Going back to $\mathcal{D}$, it follows that the modular Hamiltonian for the density matrix $\rho_B$ is the Hamiltonian which generates the image under the inverse conformal transformation of these time translations back in $\mathcal{D}$, shown on the left in figure \ref{conformal}.

To obtain the explicit expression \reef{modhcft}, we define $\zeta_\B$ to be the image of the Killing vector $2\pi R\,\partial_\tau$ under the inverse conformal transformation. This is a conformal Killing vector on the original Minkowski space which can be written as a combination of a time translation $P_t$ and a certain special conformal transformation $K_t$,
\be\labell{defzeta}
\zeta_\B = \frac{i\pi}{R}( R^2 P_t  +  K_t)
 \ee
where
 \be
i P_t = \p_t \ ,\quad{\rm and} \quad
i K_t = - [(t-t_0)^2 + |\vec{x}-\vec{x}_0|^2] \p_t - 2 (t-t_0)(x^i -
x_0^i)\p_i \ .
 \ee
It is straightforward to check that $\zeta_\B$ generates a flow which remains entirely in $\mathcal{D}$, acting as a null flow on $\p \mathcal{D}$ and vanishing on the sphere $\partial B(R,x_0)$ and at the future and past tips of $\mathcal{D}$. The generator of this flow in the underlying CFT may be written covariantly as
\be\labell{modhcft3x}
H_\B =
\int_{\cal{S}}
d\Sigma^\mu\, T_{\mu\nu}\, \zeta^\nu_\B
 \ee
where $d\Sigma^\mu$ is the volume-form on the $(d-1)$-dimensional surface $\cal S$. The integral may be evaluated on any spatial surface $\cal{S}$ within the causal diamond $\cal D$
whose boundary is $\p \B$, but for the particular choice
$\mathcal{S} = \B(R,x_0)$, we recover \reef{modhcft}. Note that the normalization of
the conformal Killing vector $\zeta_B$ was chosen in \reef{defzeta} to ensure that modular Hamiltonian $H_B$ and the Hamiltonian on the hyperbolic cylinder $H_\tau$ are
related by
$H_B=2\pi R \,U_0\,H_\tau\,U_0^{-1}$ where $U_0$ is the unitary transformation which implements
the conformal mapping between the two backgrounds \cite{chm}.

In summary, starting from the vacuum state of any conformal field theory and considering a ball-shaped region $B$, the first law (\ref{firstlaw}) simplifies to
\be
\labell{firstlaw3}
\d S_B =\d  E_B  ,
\ee
where we define
\be
\labell{defE}
 E_B \equiv 2\pi\ \int_{\B(R,x_0)}\!\!\!\!\!\!\!\! d^{d-1}x\  \frac{R^2 -
|\vec{x}-\vec{x}_0|^2}{2R}\   \langle  T_{tt}(t_0,\vec{x}) \rangle \ .
\ee

\subsection{Interpretation of the first law in holographic CFTs \labell{ss:holographic} }

The first law \reef{firstlaw} reviewed in the previous section is a general result. Hence for ball-shaped regions in an arbitrary CFT in any number of spacetime dimensions,  $\delta S_\B = \d E_\B$ with $E_B$ defined in eq.~\reef{defE}. We will be interested in understanding this relation for holographic CFT's with a \emph{classical} bulk dual, i.e. theories for which at least a subset of the states have a dual interpretation as smooth, asymptotically AdS field configurations. In this case, the vacuum state of the boundary CFT corresponds to pure anti-de Sitter space, while certain small perturbations around the vacuum state should correspond to  spacetime geometries that are small perturbations around empty AdS.\footnote{More precisely, the states that we will consider have energy of order $\e\,c_T$, where $c_T$ is the central charge of the boundary CFT (e.g. see \cite{central}) which provides a measure of the number of degrees of freedom
in the CFT. Since we consider classical gravity in the bulk,  $c_T \r \infty$; we take $\e <<1$ in order for the perturbation to be classical, but small. The first law relation also holds for quantum states in the bulk, whose CFT energy does not scale with $c_T$, but we will not consider them in this article.} In holographic theories, both $S_B$ and $E_B$ should match with observables on the gravity side, so $\delta S_\B = \d E_\B$ will translate into a constraint $\delta S^{grav}_\B = \d E^{grav}_\B$ that must be satisfied for any spacetime dual to a small perturbation of the vacuum AdS spacetime.

\subsubsection{Holographic interpretation of the entanglement entropy}

The holographic prescription for computing the entanglement entropy is not known in general, but in the known cases (e.g., \cite{rt1,timedep,aitor,xidong,Fursaev:2013fta}), it is given by extremizing a certain functional of the bulk metric over codimension-two bulk surfaces whose boundary coincides with $\p A$ in the boundary CFT. However, here we are only interested in the holographic entanglement entropy for a ball-shaped region $B$ in the CFT  when the total state is the vacuum or a small perturbation thereof. This particular case is well-understood due to the observation of \cite{chm}, reviewed in the previous section, that the vacuum density matrix for $B$ maps by a conformal transformation to a thermal state of the CFT on hyperbolic space.

\begin{figure}
\centering
\includegraphics[width=0.3\textwidth]{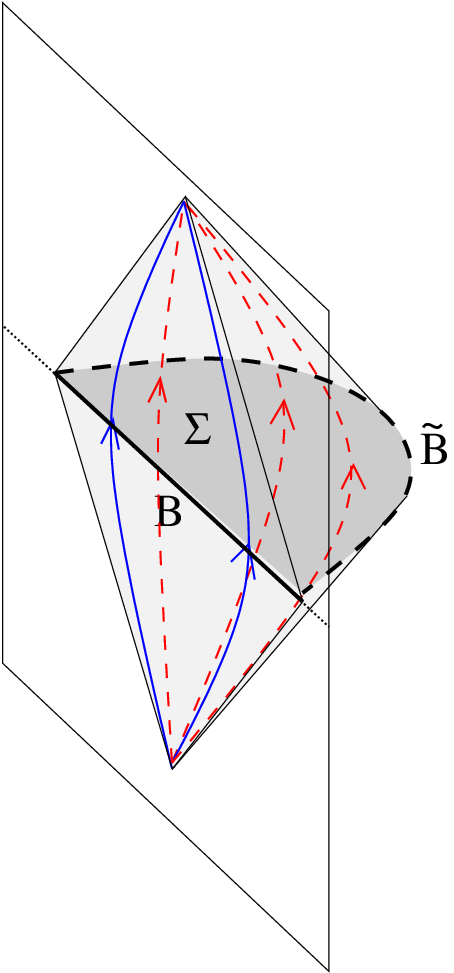}\\
\caption{AdS-Rindler patch associated with a ball $B(R, x_0)$ on a spatial slice of the boundary. Solid blue paths indicate the boundary flow associated with $H_B$ and the conformal Killing vector $\zeta$. Dashed red paths indicate the action of the Killing vector $\xi$.}
\label{hyperbolic2}
\end{figure}

Using the AdS/CFT correspondence, the thermal state of the CFT on hyperbolic space at temperature $T=\frac{1}{2\pi R}$ is dual to a hyperbolic `black hole' spacetime at this temperature, \ie AdS-Rindler space, with metric
 \be
 \label{hypmetric}
ds^2 = -\frac{\rho^2-\ell^2}{R^2} \, d \tau^2 + \frac{\ell^2\, d\rho^2}{\rho^2 -\ell^2} + \rho^2 (du^2 + \sinh^2 u\, d \Omega^2_{d-2}) \ .
 \ee
The entanglement entropy for the region $B$ equals the thermal entropy of the hyperbolic space CFT, which can be interpreted as the entropy of this
`black hole.'
In an arbitrary theory of gravity, black hole entropy is computed by evaluating the Wald functional \reef{waldent0} on the horizon. In terms of the Poincar\'e  coordinates on AdS space
\be
ds^2 = {\ell^2\over z^2} \left(dz^2 + \eta_{\mu\nu} dx^\mu dx^\nu \right) \; , \labell{poinc}
\ee
the hyperbolic `black hole' associated with the ball $B(R,x_0)$ is simply the wedge shown in figure \ref{hyperbolic2}, the intersection of the causal past and the casual future of the region $\mathcal{D}$ on the boundary. The coordinate transformation between the two metrics is described in \cite{chm}. The horizon slice approached with $\rho\to\ell$ and $\tau$ fixed in the
black hole metric \reef{hypmetric} corresponds to the hemisphere $\tilde{B} = \{t=t_0,\ (x^i - x^i_0)^2 + z^2 = R^2 \}$
in Poincar\'e coordinates. By design \cite{chm}, this is also the extremal surface in AdS bulk with boundary $\p \B$. Thus, the entanglement entropy $S_B$ for the vacuum state
can be calculated gravitationally by evaluating the Wald functional \reef{waldent0} on the surface $\tilde{B}$.

If we consider a perturbation of the original vacuum state, the perturbation of the entanglement entropy must equal the perturbation of the thermal entropy of the CFT on the hyperbolic cylinder. Assuming that this equals the perturbation to the black hole entropy, we must also have that $\delta S_\B^{grav} = \delta S_\B^{Wald}$.
In general, the entanglement entropy functional is known to differ from the Wald functional \cite{highc,xidong} by terms quadratic in the extrinsic curvature of the extremal bulk surface. These terms are important for arbitrarily-shaped entangling surfaces or general states in the CFT. However, for the special case of a spherical entangling surface considered here and a CFT in the vacuum, the extremal surface $\tilde{B}$ in the bulk is the bifurcation surface of the Killing horizon defining 
the boundary of the AdS-Rindler patch and the extrinsic curvatures of this surface vanish.
Therefore,  $\delta S_\B^{grav}$ and $\delta S_\B^{Wald}$ are equal at linear order in the perturbations we are considering.\footnote{Further, note that in the perturbed spacetime, the extremal surface will not necessarily correspond to the bifurcation surface of the AdS-Rindler horizon. However, since $\tilde \B$ is an extremal surface for the Wald functional, changes in the Wald functional due to variations in the surface come in only at second order in the metric perturbation. To calculate the Wald functional at leading order in the metric perturbation, we therefore need only evaluate $\delta S_\B^{Wald}$ on $\tilde{B}$, the bifurcation surface of the unperturbed AdS-Rindler horizon.}

To summarize, the holographic dictionary implies that
$S_\B^{grav}$ for a small perturbation around AdS is the Wald functional of the perturbed metric evaluated on $\tilde \B$.

\subsubsection{Holographic interpretation of the modular energy $E_\B$}\label{sss:moden}

In the CFT, the expression (\ref{defE}) defines $E_B$ in terms of the expectation value of the field theory stress energy tensor. On the gravity side, the latter is computed by the ``holographic stress tensor''  $T^{grav}_{\mu\nu}$, a quantity constructed locally from the asymptotic metric.\footnote{Other fields in the bulk may also contribute to the holographic stress tensor (e.g. a bilinear of gauge fields when $d=2$), but their contributions are always nonlinear in the fields and vanish at the linearized order around pure AdS that we are considering.  } For a general theory, $T_{\mu\nu}^{grav}$ can be obtained via a systematic procedure known as holographic renormalization \cite{de Haro:2000xn,Balasubramanian:1999re,Papadimitriou:2004ap}. Alternatively, as we show in sections \ref{ss:derivet} and \ref{hcurv} below,  $T^{grav}_{\mu \nu}$ can be derived using the holographic entanglement entropy function and the relation $\delta S_\B = \delta E_\B$.

The gravitational version of $E_B$ is simply obtained by replacing the stress tensor expectation value in  \eqref{modhcft3x} or (\ref{defE})  with the holographic stress tensor
\be\labell{modhcft2}
E^{grav}_\B = \int_{\cal{S}}
d\Sigma^\mu\, T_{\mu\nu}^{grav}\, \zeta^\nu_\B  =  2\pi \int_{\B(R,\vec{x}_0)}\!\!\!\!\!\!\!\! d^{d-1} x\  \frac{R^2 -
|\vec{x}-\vec{x}_0|^2}{2R}\,T^{grav}_{tt}(t_0,\vec{x})
\ee
giving $E_B^{grav}$ as an integral of a local functional of the asymptotic metric over the region $\B(R,x_0)$ at the AdS boundary.

As discussed above, $E_B$ is the conserved quantity associated with the boundary conformal Killing vector $\zeta_B$. An alternate definition \cite{iyer1} of the gravitational quantity associated with this is as the canonical conserved charge associated to translations along a bulk asymptotic Killing vector $\xi_B$ that asymptotically agrees with $\zeta_B$, $\lim_{z \r 0} \xi_B = \zeta_B$.
 We review this definition $E_B^{grav}[\xi_B]$  in section \ref{s:detail} below and show in section \ref{ss:bulkmod} that it agrees with (\ref{modhcft2}) (at least for the perturbations we are considering). Thus, for perturbations to the vacuum state, $E_B^{grav}$ can be interpreted as the perturbation to the energy of
the AdS-Rindler patch associated with the region $B$, as in figure \ref{hyperbolic2}.

Note that under the conformal map from $\mathcal{D}$ to $\mathcal{H}$, the conserved charge associated to $\zeta_B$ maps to ($2 \pi R$ times) the energy associated to $\tau$ translations, computed using either formalism.

\subsubsection{Summary}

In summary, for states of a holographic CFT with a classical gravity dual description, the CFT relation $\delta S_B = \delta E_B$ translates to a statement that the integral of the Wald functional over the bulk surface $\tilde{B}$ must equal the integral of the energy functional, as given in (\ref{modhcft2}), over the boundary surface $B$. This provides one nonlocal constraint on the metric for each ball $B$ in each Lorentz frame. The constraints may be interpreted as the statement that the perturbation to the entropy of the
AdS-Rindler patch
associated with the region $B$ equals the perturbation to the energy.

\section{The holographic first law of entanglement from the first law of black hole thermodynamics}\labell{s:forward}

In holographic CFTs, whenever the gravitational observables corresponding to $S_\B$ and $E_\B$ are known, the first law of entanglement entropy \reef{firstlaw} applied to a ball,
i.e. $\delta S_\B = \delta E_\B$, gives a prediction for the equivalence of two corresponding gravitational quantities, $\d S_\B^{grav}$ and $\d E_\B^{grav}$, in any spacetime dual to a small perturbation of the CFT vacuum state. This prediction must hold assuming the validity of the AdS/CFT correspondence and of our holographic interpretation of $E_\B$ and $S_\B$.
As we will see in the next section,
the power of this equivalence arises because in fact, we have an infinite number of predictions since $\delta S_\B = \delta E_\B$ can be applied for any
ball-shaped region in any Lorentz frame in the boundary geometry. For the case of Einstein gravity, where entanglement entropy is calculated by the Ryu-Takayanagi proposal \cite{rt1,rt2}, the equivalence of $\d S_\B^{grav}$ and $\d E_\B^{grav}$ was confirmed in \cite{relative}, and by a different method in \cite{eom}.

In this section, we will verify that $\delta S_\B^{grav} = \delta E_\B^{grav}$ follows from the equations of motion in a general theory of gravity. The crucial observation, described in the previous section, is that this gravitational relation can be interpreted as a statement of the equivalence of energy and entropy for perturbations
of AdS-Rindler space.
This equivalence follows directly from the generalized first law of black hole thermodynamics proved by Iyer and Wald \cite{iyer2}.

The Iyer-Wald theorem states that for a stationary spacetime with a bifurcate Killing horizon generated by a Killing vector $\xi$, arbitrary on-shell perturbations satisfy $\frac{\kappa}{2\pi}\delta \Sw = \delta E[\xi]$. Here $\Sw$ is the Wald entropy defined in the introduction, $E[\xi]$ is a canonical energy associated to the Killing vector $\xi$ and $\kappa$ is the surface gravity: $\xi^a \del_a\xi^b = \kappa \, \xi^b$ on the horizon.

The key observation that connects this to our holographic version of the entanglement first law is that the Iyer-Wald theorem applies to
AdS-Rindler horizons.
It is straightforward to check that the vector
\be\labell{defxi}
\xi_\B  = - \frac{2\pi}{ R}  (t-t_0) [z \p_z + (x^i-x^i_0) \p_i ] + \frac{\pi}{R} [R^2 - z^2 - (t-t_0)^2 - (\vec{x}-\vec{x_0})^2] \, \p_t
\ee
is an exact Killing vector of the standard Poincar\'e metric \eqref{poinc}, which vanishes on $\tilde{\B}(R,\vec{x}_0)$. This vector is in fact proportional to $\p_\tau$ in the AdS-Rindler coordinates \reef{hypmetric}. Thus, the hemisphere $\tilde{\B}$ is the bifurcation surface of the Killing  horizon for $\xi_B$  and the region $\Sigma(R,\vec{x}_0)$ enclosed by $\tilde{\B}$ and $\B$ is a spacelike slice that plays the role of the black hole exterior. The Iyer-Wald theorem applies, and the Killing vector has been normalized such that $\kappa = 2\pi$,  so $\delta S_\B^{Wald} = \delta E_\B[\xi_\B]$.

The definition of modular energy entering the above equality is the Iyer-Wald one; we show in section  \ref{ss:bulkmod} that this quantity agrees with $\d E_B^{grav}$ defined in terms of the holographic stress tensor. Finally, we argued in the previous section that $\delta S^{grav}_B = \delta S_\B^{Wald}$, and therefore it follows that $\delta S^{grav}_B = \delta E_B^{grav}$.
This generalizes the result of \cite{relative} to an arbitrary higher-derivative theory of gravity.

\section{Linearized gravity from the holographic first law}\labell{s:converse}

In the previous section, making use of the theorem of Iyer and Wald \cite{iyer2}, we argued that in a general theory of gravity, any perturbation to AdS satisfying the linearized gravitational equations will obey the holographic version of the entanglement first law, i.e.  $\delta S^{grav}_\B = \delta E^{grav}_\B$. In this section, we will demonstrate a converse statement: any asymptotically AdS spacetime for which $\delta S^{grav}_\B = \delta E^{grav}_\B$ for all balls $B$ in all Lorentz frames must satisfy the linearized gravitational equations and have the appropriate boundary conditions at the asymptotic boundary.

We begin in section \ref{ss:derivet} by showing that $\delta S^{grav}_\B = \delta E^{grav}_\B$ applied to infinitesimal ball-shaped regions allows us to determine the holographic stress tensor in a general theory of gravity and to constrain the asymptotic behavior of the metric. In section \ref{ss:genarg}, we explain how $\delta S_\B^{grav} = \delta E_\B^{grav}$, when applied to balls of arbitrary radius and centered at arbitrary locations in arbitrary Lorentz frames, can be used to deduce the linearized gravitational equations of motion, generalizing the results of \cite{eom}. Since we have already argued in section \ref{s:forward} that these equations of motion imply $\delta
S^{grav}_\B = \delta E^{grav}_\B$, it follows that the holographic version of the entanglement first law is {\it equivalent} to the linearized gravitational equations
in general theories of gravity.

In situations where the metric perturbation is the only field turned on in the bulk, the asymptotic behavior of the metric together with the linearized equations of motion determine the metric perturbation everywhere. In this case, knowledge of the entanglement functional allows us to recover the complete mapping from states to dual spacetimes at the linearized level.

\subsection{The holographic stress tensor from the holographic entanglement functional}\labell{ss:derivet}

To begin, we show that given the holographic prescription for computing entanglement entropy, the equation $\delta S_\B = \delta E_\B$ applied to ball-shaped regions of vanishing size can be used to determine the relation between the expectation value
of the field theory stress tensor and the asymptotic metric in the dual spacetime.

Recall the result \reef{defE},
\be
\labell{defE2}
\delta E_{\B(R,x_0)} = 2 \pi \int_{\B(R,x_0)}\!\!\!\!\!\!\!\! d^{d-1} x \  {R^2 - |\vec{x}-\vec{x}_0|^2 \over 2 R} \,  \delta \langle  T_{tt}(t_0,\vec{x}) \rangle \ .
\ee
In the limit of a very small spherical region, i.e.~$R \to 0$, the expectation value of the stress tensor is approximately constant throughout the ball $\B(R,x_0)$. Thus, the leading contribution to $\delta E_\B$ is obtained by replacing $ \delta \langle  T_{tt}(t_0,\vec{x})  \rangle$  with its central value $ \delta \langle  T_{tt}(t_0,\vec{x}_0)  \rangle \equiv \delta \langle T_{tt}(x_0)  \rangle $, which yields
\be
\delta E_{\B(R,x_0)} \xrightarrow[]{R \r 0} 2 \pi\,   \delta \langle  T_{tt}(x_0)  \rangle \int_{|x| \leq R}\!\!\!\!\!\! d^{d-1} x \  {R^2 - \vec{x}^2 \over 2 R}
= {2 \pi R^d \Omega_{d-2} \over d^2 - 1}\  \delta \langle  T_{tt}(x_0)  \rangle
\ee
where $\Omega_{d-2}$ is the volume of a unit (${d-2}$)-sphere.  Now using the CFT relation $\delta E_\B = \delta S_\B$, we find
\be
\delta \langle  T_{tt}(x_0) \rangle
= {d^2 - 1 \over 2 \pi\, \Omega_{d-2}} \lim_{R \to 0} \left( {1 \over R^{d}}\, \delta S_{\B(R,x_0)} \right)\ . \;
\labell{dictionary1}
\ee
The meaning of this equation is the following: $S_\B$ is a bulk Wald functional that depends on a small metric perturbation $h_{ab}$, as well as the radius $R$ and center $x_0$ of the entangling surface. The above equation tells us that $ S_\B[h]/R^{d}$ cannot be arbitrary, but rather it must have a finite limit as $R \r 0$. 

Repeating the same calculation for a frame of reference defined by some proper
$d$-velocity $u^\mu$, we find
\be
\labell{dictionary}
 u^\mu u^\nu \, \delta \langle  T_{\mu \nu}(x_0)\rangle  = {d^2-1 \over 2 \pi \Omega_{d-2}} \lim_{R \to 0} \left( {1 \over R^{d}}\, \delta S^{(u)}_{\B(R,x_0)} \right) \; ,
\ee
where $ \delta S^{(u)}_{\B(R,x_0)}$  is the variation of the entanglement entropy for a ball of radius R, centered at $x_0$ on a spatial slice in the frame of reference of an observer moving with the $d$-velocity $u^\mu$.
From the result (\ref{dictionary}), it is clear that given the bulk prescription for calculating $\delta S_\B$, this formula 
provides us the holographic dictionary for the stress tensor.

\subsubsection*{Example: theories with entropy equal to area}

As an example, consider a $d$-dimensional field theory for which the entanglement entropy is computed by the Ryu-Takayanagi prescription \cite{rt1} in the dual
($d+1$)-dimensional bulk
\be
S^{grav}_\B = \frac{\mathcal{A}_{\tilde{\B}}}{4 \Gn} \ .
\ee
We consider a small metric perturbation $h_{ab}$ of the AdS metric \eqref{poinc}, chosen to be in radial gauge, 
\be
h_{z\mu} = h_{zz} =0 \ .
\ee
The change in the entanglement entropy of the ball due to this bulk perturbation is
\be
\delta S^{grav}_\B = {R \ell^{d-3} \over 8 \Gn} \int_{|\vec{x}-\vec{x}_0| \leq R}\!\!\!\!\!\!\!\! d^{d-1}{x}\  z^{2-d} \, \left(\delta^{ij} - {1 \over R^2} (x^i-x^i_0)(x^j-x^j_0)\right) h_{ij} (z,t_0,\vec{x})\ . \labell{einsent}
\ee
In the limit $R \r 0$, we can replace $h_{ij}(z,x^\mu)$ by $ h_{ij} (z,x_0^\mu)$ under the integral sign. To compute the $R$-scaling of the entropy and check whether it can satisfy \eqref{dictionary1}, it is useful to define the rescaled variables
\be
\hat x^i = \frac{x^i-x^i_0}{R} \;, \;\;\;\;\; \hat z = \frac{z}{R}
\ee
which are to be kept fixed as $R \r 0$.
Then, the only way that \eqref{einsent} has a finite limit as $R \r 0$ is if
\be
h_{\mu\nu} (z,x_0^\lambda) \xrightarrow{\;\;z \r 0\;\;} z^{d-2}\ h^{(d)}_{\mu\nu}(x_0^\lambda)
\ee
where $h_{\mu\nu}^{(d)}$ does not scale with $R$. Performing the $x^i$ integral and substituting into \eqref{dictionary1}, we find
\be
 \delta \langle T_{tt} \rangle= \frac{d \ell^{d-3}}{16 \pi \Gn}\,  h^{(d) i}{}_{i} \ .
\ee
In order to generalize this result to an arbitrary Lorentz
frame as in \eqref{dictionary}, it is useful to rewrite $h^{(d) i}{}_{i} =h^{(d)}_{00} - \eta_{00}\, h^{(d)\l}_\l $. Passing to an arbitrary frame and equating the coefficients of $u^\mu u^\nu$, we find
\be
\delta \langle T_{\mu \nu} \rangle  = {d \ell^{d-3} \over 16 \pi \Gn}( h^{(d)}_{\mu \nu} - \eta_{\mu \nu}\, h^{(d)\l}_\l) \ .
\ee
Now tracelessness and conservation of the  CFT stress tensor imply that this leading perturbation of the bulk metric must  satisfy
\be\labell{hdcon}
h^{(d)\mu}_{\,\mu} =0 \;, \;\;\;\;\; \p_\mu h^{(d)\mu\nu} =0 \ .
\ee
These equations correspond to the initial value constraints on the $z=0$ surface in Einstein gravity. Applying the tracelessness condition allows the stress tensor to be simplified to
\be
\label{holoT}
\delta \langle T_{\mu \nu} \rangle \equiv \delta T_{\mu\nu}^{grav} = {d \ell^{d-3} \over 16 \pi \Gn}
\, h^{(d)}_{\mu \nu}\ \ .
\ee
Of course, this expression is the usual result for the linearized holographic stress tensor in Einstein gravity in $AdS_{d+1}$ \cite{Balasubramanian:1999re,de Haro:2000xn}.

In section \ref{ss:bulkmod} we show, using a scaling argument, that even in the presence of higher derivative terms, the CFT stress tensor $T_{\mu\nu} \propto h^{(d)}_{\mu\nu}$, but with a non-trivial coefficient that depends on the higher curvature couplings\footnote{This result can understood using conformal invariance, since $h^{(d)}_{\mu\nu}$ is the only spin-2 tensor that we can write down with scaling dimension $d$ under $z \to \lambda z$. Here we are assuming that there are no scalar fields, coupled linearly to curvature, with mass tuned so that the conformal dimension of the dual operators is $\Delta = d$, and similarly for other matter fields.} --- see also section
\ref{hcurv}.

\subsection{The linearized Fefferman-Graham expansion \labell{fg}}

We have just shown how the $R \rightarrow 0$ limit of the first law relation constrains the leading behavior of the metric for small $z$ and determines the holographic stress tensor. By equating terms at higher orders in the expansion of $\delta S^{grav}_{B(R,x_0)} = \delta E^{grav}_{B(R,x_0)}$ in powers of $R$, we can obtain additional constraints on the metric. At each higher order in $R$, the equations involve successively higher terms in the Fefferman-Graham expansion of the metric (i.e. the expansion in powers of $z$). In \cite{eom}, it was shown, for theories with holographic entanglement entropy computed by area, that these constraints completely determine the linearized metric to all orders in the Fefferman-Graham expansion. At the linearized level, this gives the complete metric perturbation everywhere in the bulk, and the result is precisely the solution to the linearized Einstein's equations with boundary behavior governed by the holographic stress tensor.

While we could apply the same approach to more general theories of gravity, we will instead take another route that leads to the full equations of motion without having to assume a series expansion for the quantities in the first law relation.

\subsection{Linearized equations from the holographic entanglement functional}\labell{ss:genarg}

\begin{figure}
\centering
\includegraphics[width=0.25\textwidth]{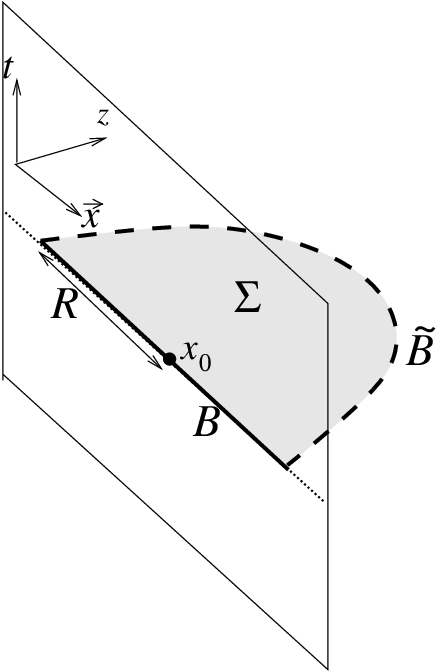}\\
\caption{Notation for regions in AdS$_{d+1}$, with radial coordinate $z$ and boundary space coordinate $\vec{x}$. $B(R,x_0)$ is the $(d-1)$-dimensional ball on the $z=0$ boundary of radius $R$ centered at $\vec{x}_0$ on the spatial slice at time $t_0$. $\tilde{B}$ is the $(d-1)$-dimensional hemispherical surface in AdS ending on $\partial B$, and $\Sigma$ is the enclosed $d$-dimensional spatial region.}
\label{notation2}
\end{figure}

In this section, we will show that knowledge of the holographic entanglement functional allows us to deduce the linearized gravitational equations for the entire dual spacetime, by making use of the relation $\delta S_\B = \delta E_\B$ for ball shaped regions $\B(R, x_0)$ of arbitrary radius $R$ and center position $x^\mu_0$ in arbitrary Lorentz frames.

Figure \ref{notation2} shows the unperturbed bulk AdS spacetime, with the region $\B(R, {x_0})$ on the boundary, together with the bulk extremal surface $\tilde{\B}(R,{x_0})$ with the same boundary as $\B$ and the spatial region $\Sigma$ on a constant time slice bounded by these two surfaces.
Using the definition (\ref{modhcft2}) and the result (\ref{dictionary}) for the holographic stress tensor, we can write the quantity $\delta E^{grav}_\B$  as an integral over the corresponding region $\B(R, {x_0})$ on the boundary of the dual spacetime of some local quantity, a  ($d-1$)-form, that is  constructed from the asymptotic limit of the metric perturbation $h_{ab}$.

In a similar way, the holographic entanglement functional gives us a prescription for writing the entanglement entropy $\delta S_\B$  as an integral over the extremal surface $\tilde{\B}(R,{x_0})$ in the bulk (shown in figure \ref{hyperbolic2}). Again, the form that we integrate is locally constructed from the metric perturbation $h_{ab}$ (and possibly matter fields). The relation $\delta S^{grav}_\B = \delta E^{grav}_\B$ then places a constraint on the perturbation: the two integrals corresponding to $\delta E^{grav}_\B$ and $\delta S^{grav}_\B$ must be equal. This must be true for any $R$ and $x^\mu_0$, in any Lorentz frame.  We will show that this infinite set of nonlocal integral equations together implies the local differential equations $\delta E^g_{ab} = 0$, where $\delta E^g_{ab}$ are the linearized  gravitational equations of motion.

\subsubsection*{Turning the nonlocal constraint into a local equation}

To convert the nonlocal integral equations into a local equation, the strategy is to make use of the machinery used by Iyer and Wald to derive the first law from the equations of motion.  The Iyer-Wald formalism is reviewed in detail in the next section, but for now we just need one fact: the crucial step in the derivation is the construction of a ($d-1$)-form $\bchi$  that satisfies
\be
\int_B \bchi = \delta E^{grav}_\B \qquad
\int_{\tilde{\B}} \bchi = \delta S^{grav}_\B  \label{chi1}
\ee
and for which $d \bchi = 0$ on shell (i.e. when the gravitational equations of motion are satisfied). The first law follows immediately by writing $\int_\Sigma d\bchi = 0$ and applying Stokes theorem (i.e. integrating by parts).

To derive local equations from the gravitational first law, we will show that there exists a form $\bchi$ which satisfies the relations  \eqref{chi1} \textit{off shell}, and whose derivative is
\be\labell{chi3}
d \bchi  =  -2\xi_\B^a \, \delta E^g_{ab}\,\bepsilon^b
\ee
where the $d$-form $\bepsilon^b$ is the natural volume form on co-dimension one surfaces
in the bulk 
(defined in eq.~\eqref{defeps}),
$\xi_\B$ is the Killing vector (\ref{defxi}) that vanishes on $\tilde \B(R,{x}_0)$,  and $\delta E^g_{ab}$ are the linearized gravitational equations of motion. In addition, we will require that %
\be
d\bchi|_{\p M} = 0\label{chi6}
\ee
where $\p M$ is the AdS boundary, assuming the tracelessness and conservation of the holographic stress tensor.\footnote{This in turn follows from the conservation and tracelessness of the CFT stress tensor.}
 This ensures that
the energy $E_B^{grav}$ does not depend on the surface $\mathcal{S}$ on the boundary that we use to evaluate it,
as long as $\p \mathcal{S} = \p B$. Note that on $\Sigma$, only the time components of $\xi_\B^a$ and $\bepsilon^b$ are non-vanishing, so only the $tt$ component of the gravitational equations appears on the right-hand side of eq.~\eqref{chi3}.

The derivation of these statements in a general theory of gravity relies on the Iyer-Wald formalism \cite{iyer2} and is deferred to section \ref{s:detail}. Now we will show that the existence of the form $\bchi$ with these properties implies the equations of motion. The relation $\delta S_B^{grav} = \delta E_B^{grav}$ gives
\be
0 = \delta S^{grav}_\B - \delta E^{grav}_\B = \int_{\tilde{\B}} \bchi - \int_{\B} \bchi = \int_{\partial \Sigma} \bchi = \int_{\Sigma} d \bchi =  -2\int_{\Sigma} \xi_\B^t\, \delta E^g_{tt}\, \bepsilon^t \ .
\ee
Multiplying this result by $R$ and then taking the derivative with respect to $R$, we obtain
\be
\int_{\tilde{\B}}  (R \xi_B^t)\, \delta E^g_{tt}\,  \hat{r} \cdot \bepsilon^t +  2\pi R \int_\Sigma  \delta E^g_{tt}\, \bepsilon^t = 0 \ .
\ee
The first term vanishes because $\xi_\B = 0$ on $\tilde{\B}$, so we find that
\be
\int_\Sigma \delta E^g_{tt}\, \bepsilon^t = 0
\ee
for any $\Sigma(R,\vec{x}_0)$.  As we show in appendix \ref{app:slide}, this implies that the integrand vanishes everywhere, i.e. $\delta E^g_{tt} = 0$, as we wished to show.

So far we have used the first law for every ball $\B(R,\vec{x}_0)$ in a spatial slice at fixed $t$.  More generally, demanding $\delta S_\B^{grav} = \delta E_\B^{grav}$ in a frame of reference defined by a $d$-velocity vector $u^\mu$ implies $u^\mu u^\nu \delta E^g_{\mu\nu} = 0$, where the index $\mu = 0, \dots, d-1$ runs over the boundary coordinates. Since this holds for any $u^\mu$, we have
\be\labell{bdeq}
\delta E^g_{\mu\nu} = 0 \ .
\ee
These are all the components of the gravitational equations of motion along the boundary directions.

To obtain the remaining equations $\delta E^g_{z\mu} = 0$ and $\delta E^g_{zz} = 0$, we appeal to the initial value formulation of gravity, in a radial slicing where these are the constraint equations. This formulation guarantees that if these constraints are satisfied at $z = 0$, and the other equations (\ref{bdeq}) hold everywhere, then the constraints hold for all $z$.\footnote{In Einstein gravity, this follows from the Bianchi identity by a standard argument \cite{Wald:1984rg}.} The vanishing of the constraints at $z=0$ follows from eq.~(\ref{chi6}) combined with eq.~\eqref{chi3}, or ultimately from the conservation and tracelessness of the holographic stress tensor.

In detail, we have using the Noether identity (discussed in appendix \ref{s:noether}) linearized about the AdS background,
\be
\label{nident}
\del_a (\delta E^g)^{ab} = 0 \ .
\ee
Using the vanishing of $E^g_{\mu\nu}$, the general solution to \eqref{nident} can be written as:
 \be
\d E^g_{z\mu} = z^{d-1} C_\mu  \,, \quad\d E^g_{zz} = z^{d-2} C_z  - \half z^{d} \partial_\mu C^\mu \,,
\label{Nsol}
 \ee
 for unfixed $C_\mu, C_z$ which are functions of the boundary coordinates. We simply
 need to show that $C_\mu, C_z$ must vanish. This is achieved
 by the requirement \eqref{chi6} which (using eq.~\eqref{chi3} and \eqref{Nsol}) gives:
 \be
 0 = \left. d \bchi \right|_{\p M} = - \left(\zeta_B^\mu C_\mu + \tilde\zeta_B^z C_z \right) dt \wedge d x^1 \ldots \wedge dx^{d-1} \; .
 \ee
 Here, we have defined $\tilde\zeta^z_B \equiv \lim_{z \rightarrow 0} (z^{-1} \xi^z_B) = -2\pi R^{-1} (t-t_0)$
 which is related to the boundary conformal Killing vector via: $\p_\mu (\zeta_B)_\nu + \p_\nu (\zeta_B)_\mu  =  2 \eta_{\mu \nu}\, \tilde\zeta^z_B$. Since it is possible to construct
 $\bchi$   for all possible boundary regions $B$ and in all Lorentz frames, it follows that $C_\mu = C_z = 0$.

In summary, we can obtain the full set of linearized gravitational equations, if we can show that a form  $\bchi$ exists, which satisfies eqs.~\eqref{chi1}, \eqref{chi3} and \eqref{chi6}. We do this in section \ref{s:detail}.

\medskip

\subsubsection*{Example: linearized Einstein equations from $S = {\cal A}/4G_N$}

In section \ref{s:detail}, we will prove that $\bchi$ exists in a general theory, but we first give the explicit formula for Einstein gravity without introducing any additional formalism. Consider the case of a holographic CFT for which the field theory entanglement entropy of a region $A$ is equal to one quarter the area of the bulk extremal-area surface with boundary $\partial A$. In this case, writing the metric perturbation as
$h_{\mu \nu} = z^{d-2} H_{\mu \nu}$, we are looking for a form $\bchi$ whose exterior derivative, restricted to $\Sigma$, is proportional to the $tt$ component of the Einstein equation, and which satisfies
\be
\int_{\B} \bchi = \delta E_B^{grav} = {d \over 16 G_N R} \int_\B d^{d-1} x\, (R^2 - |\vec{x}-\vec{x}_0|^2)\, H^{i} {}_i
\label{check1}
\ee
and
\be
\int_{\tilde{\B}} \bchi = \delta S_B^{grav} = {\ell^{d-3} \over 8 G_N R} \int_{\tilde{\B}} d^{d-1} x (R^2 H^{i} {}_i - (x-x_0)^i (x-x_0)^j H_{ij}) \; .
\label{check2}
\ee
Here, we have used eqs.~(\ref{defE}) and (\ref{holoT}) to write an
explicit expression for $\delta E_B^{grav}$, making use of (\ref{hdcon})
to replace $H_{tt}$ with $H^i{}_i = \d^{ij} H_{ij}$. The expression for $\delta S_B^{grav}$ was taken from \cite{relative,eom}.

A form $\bchi$ that satisfies the above requirements is
\be\label{covein}
\bchi = -\frac{1}{16\pi \Gn}\left[ \delta(\del^a\xi_B^b \, \bepsilon_{ab}) + \xi_B^b \, \bepsilon_{ab}(\del_c h^{ac} - \del^a h^c{}_c)\right] \
\ee
where $\bepsilon_{ab}$ is defined in eq.~\eqref{defeps}. 
The restriction of $\bchi $ to $\Sigma$ is
\bea
  \bchi|_\Sigma &=&  {z^d \over 16\pi \Gn }\left\{   \bepsilon^t {}_z \left[\left({2\pi z \over R} + {d \over z} \xi^t + \xi^t \partial_z\right) H^i{}_i\right]+ \right. \\
  & & \left.\quad\quad +  \bepsilon^t {}_i \left[ \left({2\pi (x^i-x_0^i) \over R} + \xi^t \partial^i \right) H^j{}_{j} - \left({2\pi (x^j-x_0^j) \over R} + \xi^t \partial^j \right) H^i{}_{j} \right]\right\} \notag
\eea
 where $\xi^t = {\pi \over R} (R^2 - z^2 -|\vec{x}-\vec{x}_0|^2)$. Using this expression, it is straightforward to verify eqs.~\eqref{check1} and (\ref{check2}), and also check that
\be
d \bchi |_\Sigma = -2 \xi^t \ \delta E^g_{tt}\  \bepsilon^t
\ee
where
\be
\delta E^g_{tt} = -{z^d \ell^{2-d} \over 32 \pi \Gn} \left(\partial_z^2 H^i{}_{i} + {d+1 \over z} \partial_z H^i{}_{i} +  \partial_j \partial^j H^i{}_{i} -  \partial^i \partial^j H_{ij}\right)
\ee
is the $(tt)$-component of the linearized Einstein equations.\footnote{Here, $E^g_{ab}$ is defined by varying the  action with respect to $g^{ab}$ and dividing by $\sqrt{-g}$, as usual.} Conservation $d\bchi|_{\p M} = 0$ follows from the conservation and tracelessness of the CFT stress tensor, so the other components of the Einstein equations $\delta E^g_{ab} = 0$ are also satisfied by the argument above.\footnote{For the case of Einstein gravity, we have $\delta E_{zz} \propto H^\mu {}_\mu$ and $\delta E^g_{z \mu} \propto \partial_\mu H^\mu {}_\nu - \partial_\nu H^\mu {}_\mu =0$ so the vanishing of these 
expressions at $z=0$ follows immediately from the tracelessness and conservation of the holographic stress tensor, using (\ref{holoT}).} Thus, for theories where the Ryu-Takayanagi area formula computes entanglement entropies, the non-local equations $\delta S_\B^{grav} =\delta E_\B^{grav}$ are equivalent to the linearized Einstein equations.

\section{Linearized equations in general theories of gravity}\labell{s:detail}

In this section, we review the formalism used by Iyer and Wald to prove a version of the first law of black hole thermodynamics in general theories of gravity (section \ref{ss:covariant}), and apply it in section \ref{ss:formdef} to construct a form $\bchi$ with the properties outlined in section \ref{ss:genarg}. We also argue in section \ref{ss:bulkmod} that the energy for a perturbed
AdS-Rindler spacetime
as defined by Iyer and Wald is equivalent to the energy defined using the holographic stress tensor in eq.~\reef{defE}.

\subsection{The covariant formalism for entropy and conserved charges}\labell{ss:covariant}

We begin by introducing notation and setting up the Iyer-Wald formalism \cite{wald0,iyer1}. A helpful general discussion motivating this formalism can be found in \cite{Lee:1990nz}.

\subsubsection*{Basic definitions}

Let $\bL$ be any gravitational Lagrangian, viewed as a $d+1$-form
 \be
\bL = {\cal L} \, \bepsilon \; \; ,
 \ee
where ${\cal L}$ is constructed from the metric, curvature tensors, and their covariant derivatives. Here, $\bepsilon$ is the volume form\footnote{Note that $\varepsilon_{a_1\cdots a_{d+1}}$ is an antisymmetric {\it tensor}, and our sign convention is $\bepsilon_{zti_1\cdots i_{d-1}} = + \sqrt{-g}$ .}
 \be
\bepsilon = \frac{1}{(d+1)!}\,\varepsilon_{a_1 \cdots a_{d+1}} dx^{a_1} \wedge 
\cdots \wedge dx^{a_{d+1}} \label{defvol} \ .
 \ee
For later convenience, we also define:
 \be
\bV_a = \frac{1}{d!}\,\varepsilon_{ab_2\cdots b_{d+1}}dx^{b_2}\wedge\cdots\wedge
dx^{b_{d+1}} \ , \quad \bV_{ab} = \frac{1}{(d-1)!}\,\varepsilon_{ab c_3\cdots
c_{d+1}}dx^{c_3}\wedge\cdots\wedge dx^{c_{d+1}} \label{defeps} \ .
 \ee
Denoting the dynamical fields collectively by $\phi = \{g_{\mu\nu}, \dots\}$, the
variation of $\bL$ under a general variation of the fields takes the form
 \be\labell{varL}
\delta \bL = \bE^\phi \delta \phi + d\bTheta(\delta\phi)
 \ee
where $\bE^\phi=0$ are the equations of motion for the theory, and $\bTheta$ is called the symplectic potential
current.\footnote{This potential $\bTheta$, and similarly the Noether charge form $\bQ$ below, have
ambiguities related to boundary terms in the Lagrangian and shifting by an
exact form.  Implicitly, these ambiguities, discussed in \cite{iyer1,ted1}, have been fixed
to simplify our formulae here but would not affect our arguments.} In the first term,
a sum over fields $\phi$ with indices contracted appropriately is implied.

\subsubsection*{Definition of Wald entropy from the Noether current}

For a spacetime with a bifurcate Killing horizon associated to a Killing vector $\xi$, the Wald entropy can be defined in terms of the Noether current associated with $\xi$, as we now review.

Starting with an {\it arbitrary} vector field $\xi$, the variation of the Lagrangian under a diffeomorphism generated by $\xi^\mu$ is
 \be
 \label{xivar}
\delta_\xi \bL = d(\xi \cdot \bL)
 \ee
where the dot denotes the usual inner product of $\xi^\mu$ with the form $\bL$.\footnote{That is, given an $n$-form $N
=\frac1{n!}\,N_{a_1a_2\cdots a_n}dx^{a_1}\wedge dx^{a_2}\wedge\cdots\wedge dx^{a_n}$,
$\xi\cdot N=\frac1{(n-1)!}\,\xi^b N_{b a_2\cdots a_n}dx^{a_2}\wedge\cdots\wedge dx^{a_n}$.} Since this represents a local symmetry of a Lagrangian field theory, Noether's theorem guarantees that we can associate to it a current $J^\mu[\xi]$ that is conserved when the equations of motion are satisfied. This Noether current (expressed as a d-form) is given by
 \be\labell{defj}
\bJ[\xi] = \bTheta(\delta_\xi \phi) - \xi \cdot \bL \ .
 \ee
Using eqs.~(\ref{xivar}) and (\ref{varL}), we can check that
 \be\labell{dj}
d\bJ[\xi] = -\bE^\phi \delta_\xi \phi \ ,
 \ee
so $J$ is conserved on shell as promised.

Because eq.~($\ref{dj}$) holds for all vector fields $\xi$, it follows \cite{waldlemma} that we can find a
$(d-1)$-form $\bQ$ such that
 \be\labell{defq}
\bJ[\xi] = d\bQ[\xi] \quad\quad(\mbox{on shell}) \ .
 \ee
Recalling that the Noether charge associated with the diffeomorphism $\xi$ is the integral of $J$ over a spacelike hypersurface $\Sigma$, we see that the existence of $\bQ$ (called the Noether charge form) allows us to express this charge as an integral over the boundary of $\Sigma$.

As shown in \cite{iyer1}, $\bQ$ can be written as
 \be\labell{qform}
\bQ[\xi] = \mathbf{W}_c\,\xi^c + \mathbf{X}^{cd}\del_{[c}\xi_{d]}   \ , \quad
\bX^{cd} = - E_R^{abcd}\bV_{ab} \ ,
 \ee
where $E_R^{abcd}$ is the `equation of motion' for the Riemann tensor, derived
as if it were an independent field in the Lagrangian:
 \be
E_R^{abcd} =  \frac{\delta {\cal L}}{\delta R_{abcd}}\equiv
\frac{\p {\cal L}}{\p R_{abcd}} - \del_{a_1} \frac{\p {\cal
L}}{\p \del_{a_1}\!R_{abcd}} + \cdots \ .
 \labell{grammar}
 \ee
Eq.~(\ref{defq}) only defines $\bQ$ on shell.  It is always possible to define $\bQ$ off shell so that
\be\labell{jcon}
\bJ[\xi] = d\bQ[\xi] + \xi^a \bC_a  \ ,
\ee
where $\bC_a$ are the constraint equations on a fixed-time slice. That is,
\be\labell{defc}
\bC_a = \sum_\phi\left[ \sum_{i=1}^r (E^\phi)_{c_1\cdots a \cdots c_r}^{b_1\cdots b_s} \phi^{c_1 \cdots c_i \cdots c_r}_{b_1\cdots b_s}\,\bV_{c_i}- \sum_{i=1}^s (E^\phi)_{c_1 \cdots c_r}^{b_1\cdots b_i\cdots b_s} \phi^{c_1  \cdots c_r}_{b_1\cdots a\cdots b_s}\,\bV_{b_i}\right] \ ,
 \ee
where $\phi$ is a type $(r,s)$ tensor, and the dots indicate that the indices appear
in the $i$th position. This is shown in \cite{iyer2} and reviewed in appendix
\ref{s:noether}. Note that only the equations of motion of non-scalar fields appear in $\bC_a$.

In a spacetime with a bifurcate Killing horizon, the Wald entropy \reef{waldent0} is now defined as
 \be\labell{waldent}
\Sw =  2\pi \int_{\mathcal{H}}\mathbf{X}^{cd}n_{cd} \ ,
 \ee
where $\mathcal{H}$ is the bifurcation surface and $n_{cd}$ is the binormal to
$\mathcal{H}$. This definition also applies to linearized excitations of a
stationary background. It is related to the Noether charge as follows. Let
$\xi$ be the Killing vector that generates the horizon and vanishes on
$\mathcal{H}$.  In general,  $\del_{[c}\xi_{d]} = \kappa\, n_{cd}$ on the horizon,
where $\kappa$ is the surface gravity. If we normalize $\xi$ so that $\kappa = 2\pi$,
then the Wald entropy equals the Noether charge
 \be\labell{waldq}
\Sw =  \int_{\mathcal{H}} \bQ[\xi] \ .
 \ee
On the stationary background, this agrees with (\ref{waldent}) because $\xi =
0$ on the bifurcation surface.  It was argued in \cite{wald0,iyer1} that
eqs.~(\ref{waldq}) and (\ref{waldent}) also agree for linearized excitations.

\subsubsection*{Definition of energy}

For perturbations of the background spacetime, we can define an energy canonically associated to a Killing vector $\xi$. Defining the symplectic current
 \be\labell{defomega}
\bomega(\delta_1 \phi, \delta_2 \phi) = \delta_2 \bTheta(\delta_1\phi) - \delta_1 \bTheta(\delta_2\phi) \ ,
 \ee
the Hamiltonian that generates translations along $\xi^\mu$ is obtained by
integrating $\bomega$ over a Cauchy surface $\mathcal{C}$,
 \be
\delta H_{W}[\xi] = \int_\mathcal{C}\bomega(\delta \phi, \delta_\xi \phi)  \ .
\ee
This can be rewritten using
\bea
\delta \bJ[\xi] &=& \delta \bTheta(\delta_\xi \phi) - \xi \cdot d\bTheta(\delta \phi)  \\
&=& \bomega(\delta_\xi \phi, \delta \phi) + d(\xi \cdot \bTheta(\delta \phi))\labell{dj2}
\eea
where we used the background equations of motion $E^\phi = 0$ and the formula for the Lie derivative of a form,
 \be
\delta_\xi \mathbf{u} \equiv {\cal L}_\xi \mathbf{u} = \xi \cdot d\mathbf{u} + d(\xi \cdot \mathbf{u}).
 \ee
Therefore using eqs.~(\ref{jcon}) and (\ref{dj2}), we have
\be
\delta H_{W}[\xi]  =   \delta \int_\mathcal{C} \xi^a \bC_a + \int_{\p \mathcal{C}} \left( \delta \bQ[\xi] - \xi \cdot \bTheta(\delta \phi)\right) \ . \labell{formh}
 \ee
Thus, $H$ reduces to a boundary term when the equations of motion are satisfied. We define the energy $\delta E[\xi]$ for an arbitrary (i.e. not necessarily on-shell) perturbation of the background spacetime as this contribution at the asymptotic boundary,\footnote{As discussed in \cite{iyer1}, this definition can be extended to general spacetimes with the same asymptotic behavior provided that there exists a form ${\bf B}$ such that $\delta \int_{\p \mathcal{C}} {\xi \cdot {\bf B}} = \int_{\p \mathcal{C}} \xi \cdot \bTheta$. In this case, we can define $ E[\xi]  =  \int_{\p \mathcal{C}} \left(\bQ[\xi] - \xi \cdot \bf B\right)$ .}
\be
\label{defEgrav}
\delta E[\xi]  =  \int_{\p \mathcal{C}} \left( \delta \bQ[\xi] - \xi \cdot \bTheta(\delta \phi)\right) \ .
\ee

\subsection{Definition of $\bchi$}\labell{ss:formdef}

Using the notation above, we can now define the form $\bchi$ described in section \ref{ss:holographic} as
\be\labell{defchi}
\bchi = \delta \bQ[\xi_\B] - \xi_\B \cdot \bTheta(\delta \phi) \; .
\ee
As an example, the covariant formalism is applied to Einstein gravity plus a scalar field in appendix \ref{app:einsteinscalar}, leading to (\ref{covein}). In the rest of this section, we will demonstrate that $\bchi$ obeys the equations (\ref{chi1}) and (\ref{chi3}) that were needed to derive the equations of motion from $\delta S_\B^{grav} = \delta E_\B^{grav}$. That $\bchi$ also obeys eq.~\eqref{chi6} will be shown in the next section. We emphasize that it is the \textit{existence} of a form $\bchi$ with these properties that guarantees that the linearized equations of motion are equivalent to $\delta S_\B^{grav} = \delta E_\B^{grav}$. Thus, starting from only the entropy functional $\Sw$, it should be possible to recover the linearized equations even if we do not know $\bchi$, $\bL$, $\bTheta$, or $\bQ$.

The first property in eq.~\eqref{chi1} follows directly from our definition \reef{defEgrav} and the equivalence of $\delta E[\xi_B]$ and $\delta E^{grav}$, to be shown in the next section. The second property follows from
\be
\delta S^{grav}_\B = \delta \Sw = \int_{\tilde \B} \delta \bQ[\xi_B] = \int_{\tilde{\B}}\bchi \  .
\ee
The first equality was discussed in section \ref{ss:holographic}, the third equality follows from the definition of $\chi$ since $\xi_B$ vanishes on $\tilde{B}$, and the second equality was proved in \cite{iyer2}, as discussed in the previous section. The proof in \cite{iyer2} does not use the equations of motion, so it holds off shell if we define the entropy as in eq.~(\ref{waldent}).

To show eq.~(\ref{chi3}) note that for $\xi$ a Killing vector of the background, eq.~(\ref{dj2}) implies $\delta \bJ[\xi] = d(\xi \cdot \bTheta)$. Therefore
\be
d \bchi = \delta (d\bQ[{\xi_\B}] - \bJ[{\xi_\B}])
 = - \xi^a_\B \, \delta \bC_a
 = -2\xi^a_\B\, \delta E^g_{ab} \,\bV^b
\ee
where $E^g_{ab}$ is the equation of motion derived by varying the action with respect to $g^{ab}$. Note that fields vanishing on the background do not contribute to the first variation of $\bC_a$, which is why our derivation always gives the gravitational equations rather than the some combination of gravitational and matter equations.

\subsection{Equivalence of the holographic and the canonical modular  energy}\labell{ss:bulkmod}

For arbitrary perturbations to AdS, we now have two definitions of modular energy associated to a given boundary region $B$:
the ``canonical'' energy \reef{defEgrav}
\be\labell{ewald}
\delta E^{grav}_{(1)} \equiv \delta E[\xi_B] \equiv \int_{\B}\left( \delta \bQ[\xi_\B] - \xi_\B \cdot \bTheta(\delta \phi)\right)
\ee
associated with the bulk Killing vector $\xi_B$ that asymptotes to $\zeta_B$, and the ``holographic'' energy (\ref{defE})
\be\labell{eus}
\delta E^{grav}_{(2)} = \int_\B d\Sigma^\mu\  \delta T_{\mu\nu}^{grav} \, \zeta^\nu_\B
\ee
defined in terms of the holographic stress tensor (\ref{holoT}).

In order to complete our story, we must show that these two definitions of energy agree, without assuming the equations of motion. This is not to say that the formulae agree for arbitrary $h_{\mu \nu}$ -- they do not. However, equivalence follows from the restrictions on the asymptotic metric implied by $\delta S_B^{grav} = \delta E_{(2)}^{grav}$. Note that the purely gravitational result of this paper -- that the linearized equations of motion are equivalent to $\delta S_B^{grav} = \delta E_{(1)}^{grav}$  -- does not require the results of this subsection; it is needed only to map the CFT problem to the gravity problem.

Consistency of the AdS/CFT dictionary requires the two definitions to agree, since both should equal the CFT energy. This is confirmed for Einstein gravity in \cite{Papadimitriou:2005ii,Hollands:2005wt}, and rather generally in \cite{Hollands:2005ya}, but these discussions rely on the equations of motion.  Here we will demonstrate the equivalence explicitly, at the linearized level, in a way that makes clear that we do not need to start from the equations of motion.  For simplicity, we assume in this calculation that matter fields are not coupled to curvature, so only the metric appears in the linearized energy.

As we discussed in section \ref{ss:derivet} and will show in more detail in section \ref{hcurv}, metric perturbations for which the first law is satisfied at leading order in the $R$ expansion behave near the boundary as
\be
\label{pge}
h_{\mu \nu} = z^{d-2} h^{(d)}_{\mu \nu} + \dots
\ee
where the dots indicate terms at higher order in $z$. Other fall-offs, with $\delta S_B^{grav} = 0$, are also allowed; these are addressed below. The holographic stress tensor is proportional to $h^{(d)}_{\mu\nu}$, so
\be
 \d E^{grav}_{(2)} = C_2 \int_\B d^{d-1}{x}\ u^\mu\, h^{(d)}_{\mu\nu}\, \zeta_\B^\nu \ .
\ee
On the other hand, if we plug in the asymptotic expansion \eqref{pge} into eq.~\eqref{ewald}, we find
\be
\labell{de1}
\delta E_{(1)}^{grav} = C_1\int_\B d^{d-1} x \, u^\mu\, h^{(d)}_{\mu\nu}\, \zeta_\B^\nu
\ee
for some coefficient $C_1$, as shown in appendix \ref{app:formcharge}. One should in principle be able to verify that $C_1 =C_2$ by an explicit computation. However, since this can be a bit tedious in an arbitrary higher derivative gravity theory, we present below a slightly indirect but simpler argument that the two constants must be the same.

To show that the coefficients $C_1$ and $C_2$ are  equal, we use the fact that, according to our definition, the entanglement first law is
\be
\d S_\B^{grav} = \d E^{grav}_{(2)} \labell{flagain} \ .
\ee
What we have shown in the previous section is that
\be
\d E_{(1)}^{grav} - \d S_\B^{grav} = 2 \int_\Sigma \xi^a_\B\, \delta E^g_{ab} \,\bV^b \ . \labell{flwaldagain}
\ee
Plugging eq.~\eqref{flagain} into eq.~\eqref{flwaldagain}, we find
\be\label{diffee}
\d E_{(1)}^{grav} - \d E^{grav}_{(2)} =   \left( C_1 - C_2 \right) \int_\B d^{d-1} x \,  h^{(d)}_{tt}\, \zeta_\B^t =  2\ell^{d-1} \int_\Sigma  dz \, d^{d-1} x  \, z^{1-d} \delta E^g_{tt} \, \xi^t_B
\ee
As $R \r 0$, the middle term in this equation is proportional to $(C_1-C_2) R^d$. On the other hand, the term on the right-hand side starts at $O(R^{d+2})$, because $h^{(d)}_{\mu\nu}$ satisfies the linearized equations of motion at leading order, as can be explicitly checked. Thus
\be
\d E_{(1)}^{grav} = \d E^{grav}_{(2)}
\ee
for these modes.

As we discuss in section \ref{fgterms}, the other fall-offs allowed by first law are those for which $\delta S_B^{grav} = 0$ as $R \to 0$. These behave near the boundary as
\be
\label{ntexp}
h_{\mu\nu} = z^{\Delta-2}h_{\mu\nu}^\Delta + \cdots
\ee
for particular values of $\Delta > d-2 $, given in eq.~\eqref{dtds}.  These modes do not appear in the holographic stress tensor, so do not contribute to $\delta E_{(2)}^{grav}$. Their contribution to $\d E_{(1)}^{grav}$ is proportional to $z^{\Delta -d}$ as $z \r 0$, and thus vanishes  if
 $\Delta > d$.
  This means that our entire analysis applies to modes with $\Delta > d$, so the linearized equations of motion hold everywhere. Furthermore, since $\Delta$ depends on the parameters in the Lagrangian, and $\delta E_{(1)}^{grav}$ must also depend smoothly on these parameters, this conclusion also applies to modes with $\Delta < d$.  Alternatively, it can be checked explicitly that such modes obey the leading equations of motion near the boundary (there is a single term to check because $\Delta > d-2$) so eq.~(\ref{diffee}) implies $\d E_{(1)}^{grav} = \d E^{grav}_{(2)}$.

Applying this discussion in an arbitrary frame, we have now established that, at the boundary, $\bchi$ is equal to the conserved current that appears in the modular energy:
\be
\bchi|_{\p M} = d\Sigma^\mu\, T_{\mu\nu}^{grav}\,\zeta^\nu \ .
\ee
Conservation and traceless of the CFT stress tensor therefore imply $d\bchi|_{\p M} = 0$, completing the derivation in section \ref{s:converse}.

\section{Application: the holographic dictionary in higher curvature gravity \labell{hcurv}}

In section \ref{ss:derivet}, we have argued that in the limit $R \r 0$, the entanglement first law, together with the holographic entanglement functional, yields the holographic dictionary for the stress tensor. As a concrete and non-trivial application of this observation, in this section we derive the holographic dictionary for the case when the entanglement entropy is given
by a Wald functional that is polynomial in the Riemann tensor --- in other words, for a higher derivative gravity theory whose action is constructed from arbitrary powers of the Riemann tensor, but no derivatives thereof. The analysis including derivatives of the Riemann tensor is similar --- and straightforward in any particular example --- but we leave it to future work. As we will show, in this case the entanglement first law allows one to derive holographic dictionary not only for the stress tensor, but also for the other operators that couple to the metric in the context of higher derivative gravity.

The usual procedure for finding the holographic dictionary for an arbitrary gravitational theory in AdS is holographic renormalization \cite{de Haro:2000xn,Balasubramanian:1999re,Papadimitriou:2004ap}. This technique provides full information about the holographic dictionary, allowing both arbitrary sources and expectation values. One can thus  compute, in principle at least, any desired correlator of the stress tensor and the other operators that couple to classical fields in the bulk. Nevertheless, computations that use this method can become extremely tedious in the context of higher derivative gravity. The reason is that a necessary first step in holographic renormalization is to render the variational principle at the spacetime boundary well-defined, and this can be rather difficult in a general higher derivative gravity theory (e.g. \cite{Hohm:2010jc,Hollands:2005ya}).

However, if one is only interested in computing the expectation value of the stress tensor in higher derivative gravity rather than its general correlation functions,  the ``entanglement first law'' method for deriving the holographic dictionary can provide an easy alternative. The reason for this simplification --- besides not having to deal with the variational principle --- is that one can perform all calculations at linearized level, where all higher derivative gravity theories effectively reduce to $R^2$ theories. A scaling argument can then be used to argue that the linearized  answer holds quite generally.

\subsection{General results}\label{ss:generalT}

We begin with the general result derived in section \ref{ss:derivet}
\be
\delta T^{grav}_{tt}(x_0)
= {d^2 - 1 \over 2 \pi \Omega_{d-2}} \lim_{R \to 0} \left( {1 \over R^{d}}\, \delta S^{grav}_{\B(R,x_0)} \right) \ .
\labell{dictionary1a}
\ee
For a general theory of gravity, we have  
\bea
\delta S^{grav}_B = \delta S^{Wald}_B &=& \delta \left( - 2\pi \int_{\tilde{B}}E_R^{abcd} \bV_{ab}\, n_{cd} \right) \nonumber\\
&=& -2 \pi \int_{\tilde{\B}} \left( \delta E_R^{abcd} \bV_{ab}\,  n_{cd} + E_R^{abcd} \delta \bV_{ab} \, n_{cd} + E_R^{abcd} \bV_{ab} \, \delta n_{cd} \right)
\label{Xvar}
\eea
The binormal $n_{cd}$ is defined as
\be
n_{cd} = n^1_a n^2_b - n^2_a n^1_b
\ee
where $n^1$ and $n^2$ are unit vectors normal to each other and to
the bifurcation surface $\tilde B$. To linearized order in the perturbation, they are given by

\be
n^1_adx^a = - \frac{\ell}{z} \left(1- \frac{z^2}{2\ell^2} \, h_{tt} \right)  dt \;, \;\;\;\;\; n^2_adx^a = \frac{x^A \ell}{R \, z} \left(1 + \frac{z^2 }{2\ell^2 R^2} \, h_{ij}\, x^i x^j \right)  d x^A
\ee
where $x^A= x_A =\{x^i,z\}$.

Next, from the definition of $\bV_{ab}$ in eq.~\reef{defeps}, we have
\be
\delta \bV_{ab} = {1 \over 2} \, h \,  \bV_{ab} \; , \;\;\;\;\; h \equiv g^{cd} h_{cd} \ .
\ee
Substituting all these expressions into eq.~(\ref{Xvar}), we find  
\bea
\delta S^{Wald}_B = \frac{ 4 \pi \ell^{d+1}}{R} \int_{B} {d^{d-1} x \over  z^{d+2} } \, x_{A} x_B  \left[- {2 \ell^2 \over z^2 } \delta E_R^{tAtB} +  E_R^{t A t B} \left(2  h_{tt}-  h_{ij} \d^{ij} - h_{ij}\, \frac{ x^i x^j}{R^2} \right) \right]
\label{Xvar2}
\eea
For a general Lagrangian built from curvatures but no covariant derivatives of curvatures, $E_R^{abcd}$ is a function of $g^{ab}$ and $R^{abcd}$. Evaluated on an AdS background, which is maximally symmetric and thus satisfies
\be
\label{Rsym}
 R_{abcd} = - \frac{1}{\ell^2} \left( g_{ac} \, g_{bd} -  g_{ad} \, g_{bc}\right)\ ,
\ee
the Wald functional takes the following simple form

\be\label{defonec}
E_R^{abcd} = c_1  \, g^{\langle a b} g^{c d \rangle}
\ee
for some constant $c_1$.
The indices inside the $\langle \; , \rangle$ brackets are (anti)symmetrized so that the resulting object has the same symmetries as the Riemann tensor, as in \cite{Hohm:2010jc}.

To compute $\d E_R^{abcd}$, we use the chain rule

\be
\label{Eexp}
\delta E_R^{abcd} = {\partial E_R^{abcd} \over \partial g^{ef}}\, \d g^{ef} + {\partial E_R^{abcd} \over \partial R_{efgh} }\, \delta R_{efgh} \ .
\ee
The partial derivatives above are to be evaluated on the background AdS, so using eq.~\eqref{Rsym}, they can be expressed entirely in terms of products of the unperturbed AdS metric, with various contractions and symmetrizations.
Letting
\be
\d g^{ab} = - h^{ab} \;, \;\;\;\;\; \d R_{abcd} = \mathcal{R}_{abcd}
\ee
the general form of the linearized $\d E_R^{abcd}$ is then
\be
\d E_R^{abcd} = - c_2 \, g^{\langle a b} g^{c d \rangle} \, h - c_3  \, h^{\langle a b} g^{c d \rangle} + \ell^2 c_4  \, g^{\langle a b} g^{c d \rangle} \, \mathcal{R} + \ell^2 c_5  \, \mathcal{R}^{\langle a b} g^{c d \rangle} + \ell^2 c_6 \, \mathcal{R}^{abcd} \label{defCoeff}
\ee
where the first two terms come from the partial derivative of $E_R^{abcd}$ with respect to $g^{ef}$, and the last three from the partial derivative with respect to the Riemann tensor, evaluated on AdS. All indices are raised and contracted with the background metric $g^{ab}$. Note that not all coefficients $c_i$ introduced above are independent, but rather they satisfy
\be
c_2= -2 d\,  c_4  - c_5 \;, \;\;\;\;\; c_3 = 2\, c_1 - (d-1)\, c_5 -4 \, c_6 \ .
\labell{constraint}
\ee
These constraints follow from the fact that the most general  Wald functional that is linear in the Riemann tensor takes the form \eqref{ER} --- see below --- which is parametrized by just four  constants. Eq.~\eqref{relca} then shows that the six
coefficients $c_i$ satisfy two additional relations.

To finalize our computation of the linearized Wald functional, we only need to evaluate the linearized Riemann tensor, given by
\be
\mathcal{R}_{abcd} = \half (\nabla_c \nabla_b h_{ad} - \nabla_b \nabla_b h_{ac} + \nabla_b \nabla_a h_{bc} - \nabla_c \nabla_a h_{bd}) + \half ( R_{a e cd} h^e{}_b +  R^e{}_{bcd} h_{ae}) \labell{linriem} \ .
\ee
For the computation of the holographic stress tensor \eqref{dictionary1a}, we only need the leading behaviour of the Riemann tensor as $R \r 0$. This can be easily evaluated  by noting that $z \propto R$ and  $z$-derivatives of the metric perturbation dominate  over $x^\mu$-derivatives as $z \r 0$.
More explicitly, near the boundary we can  write
\be
h_{\mu\nu}(z,x^\lambda) =z^{\Delta-2} h^{(\Delta)}_{\mu\nu} (x^\l) + \cdots
\labell{metpert}
\ee
for some $\Delta$ to be determined, where the dots indicate terms at higher order in $z$. Then, we can  replace $\p_z h_{\mu\nu} = (\D-2) z^{-1} h_{\mu\nu}$ and  ignore all $x^\mu$-derivatives, because $\p_\mu \sim \mathcal{O}(1)$ will always be subleading in the $R$ expansion, as compared to $\p_z \sim \mathcal{O}(R^{-1})$. Consequently, in taking $R \r 0$, we can approximate\footnote{It is not hard to see that if we had also allowed derivatives of the Riemann tensor into the Wald functional, their linearized leading contribution to the entropy as $R \r 0$ would also be linear in $h_{\mu\nu}$ with no derivatives, due to the above scaling argument. Their contribution  would typically be of the same order as the polynomial one, and  straightforward if a bit tedious to compute.  }
\be
\label{dR1}
\left. \mathcal{R}_{\mu\nu\rho\s} \right|_{R \r 0} = \frac{\D -2}{2\ell^2} (h_{\mu\rho}  g_{\nu\s} + h_{\nu\s}  g_{\mu\rho} - h_{\mu\s}  g_{\nu\rho} - h_{\nu\rho} g_{\mu\s})
\ee
and
\be
\label{dR2}
\left. \mathcal{R}_{\mu z \nu z} \right|_{R\r 0} = \frac{1}{2z^2} [2(\D-1) - \D^2] \, h_{\mu\nu} \ .
\ee
We can then substitute this simplified expression into\footnote{The explicit expression for eq.~\eqref{defCoeff} is \tiny
\bea
\d E^{\mu\nu\rho\s}_R &= &\left[(\D d - 2 d - \D^2 + \D)\,c_4  + \frac{c_5}{2} (\D-2)-c_2 \right] h \, g^{\langle \mu\nu} g^{\rho\s \rangle}+ \left[ 2  (\D-2) c_6 - c_3 + \frac{c_5}{2} (\D d-2 d +2 -\D^2) \right] h^{\langle \mu\nu} g^{\rho\s \rangle} \nonumber \\
\d E^{\mu z\rho z}_R  &=&\left[ \frac{c_6}{2} (2\D -2 - \D^2) + \frac{c_5}{8} (\D d - 2 d +2 -\D^2)  - \frac{c_3}{4} \right] h^{\mu\nu} g^{zz}  + \left[ \frac{c_4}{2} (\D d -2 d -\D^2 +\D) + \frac{c_5}{8} (3\D -4 -\D^2)  - \frac{c_2}{2} \right] \, h \, g^{\mu\nu} g^{zz}\nonumber
\eea
 } eq.~\eqref{defCoeff} and further into eq.~\eqref{Xvar2}. Upon contracting with $x_A \, x_B$, the integrand will contain terms proportional to $h_{tt}$, $ \d^{ij} h_{ij} $ and $h_{ij} x^i x^j$;  using spherical symmetry,  the latter can be replaced by $ \vec{x}^2\,\d^{ij} h_{ij}/(d-1)$. Furthermore, we can write $h_{ij} \d^{ij} = h_{\mu\nu} \eta^{\mu\nu} + h_{tt}$. The final answer takes the form
\be
\delta S^{Wald}_B = \frac{ 4 \pi \ell^{d-3}}{R} \int_{B} {d^{d-1} x \over  z^{d-2} } \,  (A \, h_{tt} + B\, \eta^{\mu\nu} h_{\mu\nu} ) \label{dsw}
\ee
where the coefficients $A$ and $B$ are given by
\bea
A &=& \left( \frac{|\vec{x}|^2}{d-1} -R^2 \right) \left[\frac{c_1}{2} -\frac{c_3}{2} + \frac{c_5}{4}(2-2d+\D d-\D^2) +  (\D^2-d \D^2 + d \D -2) c_6 \right]\notag\\
& & \hspace{2cm} - R^2 \Delta (\Delta-1)(d-2) c_6 \nonumber \\
B&=& \frac{ |\vec{x}|^2}{d-1}\left[\frac{c_1}{2}- \frac{c_3}{2}+\frac{c_5}{4}  (2-2d+ \D-2\D^2 + d \D^2)  + c_6 (\Delta-2) \right] + \nonumber \\ &&\hspace{1.5 cm} +R^2 \left[\frac{c_1}{2}- c_2 + c_4 (\D +d \D -2d -\D^2) +\frac{c_5}{4} (3\D-4-\D^2)
\right]\ .
\eea
Using eq.~\eqref{metpert}, it is not hard to verify that the leading contribution in eq.~\eqref{dsw} scales as $R^\D$, so we must choose $\Delta = d$ to obtain a finite result in eq.~(\ref{dictionary1a}).  Thus, we find again, as in Einstein gravity, that in order for the first law of entanglement to be satisfied, the asymptotic expansion of the metric should start at order $z^{d-2}$.
Performing the integral in eq.~\eqref{dsw} with $\D=d$, we find  that
\be
\delta T^{grav}_{tt} = \alpha\, h^{(d)}_{tt} + \beta \, \eta_{tt} \, h^{(d)}_\mu {}^\mu
\ee
where the indices on $h^{(d)}$ are now raised with $\eta^{\mu\nu}$, and the two coeffcients are given by
\be
\alpha= d (- c_1 + c_3 + (d-1) c_5 + 2 d c_6 )\, \ell^{d-3} \ee
\be
\beta = [ - (d+2)c_1 + 2(d+1)c_2 + c_3 + 2 d (d+1) c_4 + (d+1) c_5 - 2(d-2) c_6  ]\, \ell^{d-3}
\ee
Generalizing the calculation to an arbitrary Lorentz frame as in section (\ref{ss:derivet}), we conclude that
\be
\delta T^{grav}_{\mu \nu} = \alpha h^{(d)}_{\mu \nu} + \beta\,  \eta_{\mu \nu} h^{(d)}_\alpha {}^\alpha
\ee
As in Einstein gravity, tracelessness and conservation of $T_{\mu \nu}$ imply that\footnote{When $\a + \b \, d =0$, the vanishing of the trace of the stress tensor no longer implies $ h^{(d)\mu}{}_\mu = 0$. Using our results from section \ref{fgterms}, it is easy to check that precisely at this value of the $c_i$, the additional scalar operator present in higher curvature gravity --- which couples to the trace of the metric --- has dimension $\D =d$, and thus appears at the same order in the asymptotic $z$ expansion as the traceless mode that couples to the CFT stress tensor.}
\be
h^{(d)\mu}{}_\mu =0 \;, \;\;\;\;\; \p^\mu h^{(d)}_{ \mu\nu}  = 0
\ee
so we have
\bea
\label{Tgeneral}
\delta T^{grav}_{\mu \nu} &=& d\, \ell^{d-3} [- c_1 + c_3 + (d-1) \, c_5 + 2 d\, c_6 ]\, h^{(d)}_{\mu \nu}\notag\\
&=& d \ell^{d-3}[ c_1 + 2(d-2)c_6]\, h^{(d)}_{\mu \nu} \ .
\eea
This gives the holographic stress for a theory in which the Wald entropy is an arbitrary function of the Riemann tensor, but not its covariant derivatives. The coefficients $c_i$ are defined in eqs.~\eqref{defonec} and \eqref{defCoeff}.

Note that to this point, we have only been considering the leading contribution to the expectation value of the stress tensor.
That is, as noted in footnote \ref{footy}, we are considering a one-parameter family of states $|\Psi(\lambda) \rangle$ with $|\Psi(0) \rangle = |0 \rangle$ and within this family, $\delta \langle T_{\mu \nu}\rangle \equiv \partial_\lambda \langle T_{\mu \nu}\rangle|_{\lambda = 0}$. However, we will now argue that our result extends beyond this leading order to give
a general prescription for $\langle T_{\mu\nu}\rangle$. In particular, the fact that $\langle T_{\mu\nu}\rangle \propto h^{(d)}_{\mu\nu}$ simply follows from conformal invariance: there is no other field in spacetime that has the correct tensor structure and transformation properties under rescalings.\footnote{There are a few exceptions to this, such as a gauge field in three space-time dimensions, which can contribute to the stress tensor at quadratic order, or when fields have finely-tuned dimensions that can add up to $d$.} Thus, the above expression for the stress tensor holds even when $h^{(d)}_{\mu\nu}$ is finite. Another way to see this fact is to note that since the theory is conformal, the only dimensionless number that characterizes the perturbation is $\varepsilon = c_T^{-1} \langle T_{\mu\nu} \rangle R^d$ in the CFT, or $h_{\mu\nu}^{(d)} R^d$ in spacetime.  Applicability of the first law only requires that $\varepsilon <<1$, see also the appendix of \cite{Blanco:2013joa}. Thus, we can either have $\langle T_{tt} \rangle$ small and $R$ finite, or $\langle T_{tt} \rangle$ finite and $R \r 0$. In the first case, we can derive the linearized gravitational equations in the entire bulk, by taking the amplitude of the perturbation to be small and using the Wald functional method. In the second case, we can derive the leading asymptotic expansion of the metric (as $z \r 0$)  for a general non-linear solution.

\subsection{Examples}

We now give some explicit examples employing the general formula (\ref{Tgeneral}) and compare with known results in the literature.

\subsubsection{The holographic stress tensor in $R^2$ gravity \labell{holotrsq}}

To begin, consider the case of an arbitrary $R^2$ gravity theory in $d+1$ dimensions, which contains all possible contractions of the Riemann tensor but no derivatives thereof. It is convenient to write the most general Lagrangian of such a theory as
\be
\mathcal{L} =  \frac{1}{16\pi\Gn}\left[ \frac{d(d-1)}{\tl^2}+ 
R + a_1\tl^2\, R_{abcd} R^{abcd} + a_2\tl^2\, R_{ab} R^{ab} + a_3\tl^2\, R^2 \right] \labell{lagrrsq} \ ,
\ee
where $\tl$ is the scale parametrizing the (negative) cosmological constant. We also use $\tl$ to set the scale in the 
curvature-squared terms, which leaves $a_i$ as dimensionless couplings controlling the strength of these
interactions. 
We assume that the parameters are chosen such that the theory admits an $AdS_{d+1}$ vacuum solution of radius $\ell$. 
In fact, it is straightforward to show the AdS radius is determined by the parameters in the Lagrangian \reef{lagrrsq} by
the following quadratic equation
\be
\frac{\ell^4}{\tl^4}- 
\frac{\ell^2}{\tl^2}+\frac{d-3}{d-1}\left(2\,a_1+d\,a_2+d(d+1)\,a_3\right)=0\ .
\labell{scales}
\ee
Of  course, $\ell=\tl$ when the $a_i$ are set to zero.
To construct the Wald entropy \reef{waldent0}, we consider the variation of the Lagrangian with respect to
the curvature, as in eq.~\reef{grammar}
\bea
 E_R^{abcd} &=& \frac{1}{16\pi\Gn}\left[\left(\frac12 
+  a_3\tl^2\, R\right) \,(g^{ac} g^{bd} - g^{ad} g^{bc}) 
\right.\labell{ER}\\
&&\quad\left. + 2 a_1\tl^2\, R^{abcd}  + \frac{1}{2}a_2\tl^2 \, \left(R^{ac} g^{bd} - R^{bc} g^{ad} - R^{ad} g^{bc} + R^{bd} g^{ac} \right)\right] \nonumber
\eea
The coefficients $c_i$ defined in eq.~(\ref{defCoeff}) are given by
\begin{align}
c_1&= \frac{1}{16\pi\Gn}\left[1 
- 2\left(2 a_1 + d a_2 +  d(d+1) a_3\right) {\tl^2\over \ell^2}\right] \ ,\quad\quad
 c_2 =  - \frac{2 (a_2 + 2 d\,a_3)}{16\pi\Gn}\,\frac{\tl^2}{\ell^2} \ ,\nonumber \\
c_3 &= \frac{1}{8\pi\Gn}\left[1 
- \left(8 a_1  + (3d-1)a_2 + 2 d(d+1) a_3\right){\tl^2 \over \ell^2}\right] \ ,
\label{relca}\\
 c_4 &= \frac{ a_3}{8\pi\Gn}\,\frac{\tl^2}{\ell^2} \;,\qquad\qquad
 c_5 = \frac{ a_2}{8\pi\Gn}\,\frac{\tl^2}{\ell^2} \;,\qquad\qquad
 c_6 = \frac{ a_1}{8\pi\Gn}\,\frac{\tl^2}{\ell^2}\ , \nonumber
\end{align}
which one can verify satisfy the constraints in eq.~\reef{constraint}.
Hence our general expression (\ref{Tgeneral}) gives
\be
 \langle T_{\mu\nu} \rangle = \frac{d \, \ell^{d-3}}{16\pi\Gn} \, 
\left[1 
+2\left( 2(d-3)a_1  -d \,a_2 -d(d+1) a_3 \right)\frac{\tl^2}{\ell^2}\right]  h_{\mu\nu}^{(d)} \labell{holostressT}
\ee
We have checked that eq.~\eqref{holostressT}  agrees perfectly with previous results in the literature that used more standard holographic techniques:  see, for example, equation (51) of \cite{Smolic:2013gz} for the case $d=3$. We have also checked that in general $d$
our answer agrees with the  holographic stress tensor of \cite{Hohm:2010jc}, when the results of that paper are applied to a flat boundary metric and the volume divergences are subtracted. Note that the covariant expression of \cite{Hohm:2010jc}  for the holographic stress tensor in terms of induced fields at the boundary obscures somewhat
 the simplicity of the final answer  \eqref{holostressT} for $\langle T_{\mu\nu} \rangle$, which is  dictated by scaling.\footnote{This scaling property might be more obvious if one used instead the Hamiltonian method for holographic renormalization \cite{Papadimitriou:2004ap}. Nevertheless, one would still need to deal with the variational principle
with that approach.}

\subsubsection{An $R^4$ example \labell{r4ex}}

As an example where higher powers of curvature appear, consider the theory
\be
I= \frac{1}{16\pi \Gn} \int d^{d+1} x \sqrt{-g} \left[ \frac{d(d-1)}{\tl^2}+R + \a\tl^6\, (R_{\mu\nu\rho\s} R^{\mu\nu\rho\s})^2\right] \; .
\ee
This particular example has been  studied previously in  section 3.4 of \cite{Smolic:2013gz}, for the case $d=3$. The authors of that paper were investigating black hole thermodynamics in the above theory, and found that in order for the first law to hold, the mass of the black hole had to be independent of the coefficient of the $R^4$ term. In this subsection, we will use the holographic entanglement method for computing the stress tensor expectation value to confirm their result.

The Wald functional for this theory reads
\be
E_R^{abcd} = \frac{1}{16\pi \Gn}\left[\half (g^{ac} g^{bd} - g^{ad} g^{bc}) + 4 \a\tl^6\, R^{abcd}  (R_{\a\b\g\d} R^{\a\b\g\d})\right]   \labell{nlw}
\ee
The four independent coefficients $c_i$ are given by
\be
c_1 =  \frac{1}{16\pi \Gn} \left(1-16 d (d+1) \alpha{\tl^6 \over \ell^6}\right)\;, \;\;\;\;\;
c_4 =\frac{2\a}{\pi \Gn} \frac{\tl^6}{\ell^6} \;, \;\;\;\;\; c_5=0\;, \;\;\;\;\;c_6= \frac{d(d+1)\a}{2\pi \Gn} {\tl^6 \over \ell^6}
\ee
so our general expression (\ref{Tgeneral}) gives
\be
\langle T_{\mu \nu} \rangle = \frac{d \ell^{d-3}}{16 \pi \Gn} \left(1 + 16d (d+1) (d-3) \a
\frac{\tl^6}{\ell^6} \right) h_{\mu \nu}^{(d)}
\ee
Thus, precisely in $d=3$ we have $\langle T_{tt} \rangle = 3 h^{(3)}_{tt}/(16\pi\Gn)$. The explicit solution (142)-(143)  in \cite{Smolic:2013gz} for the metric of the black hole in presence of the $R^4$ term shows that $h_{00} = m$ is uncorrected by the higher derivative term. Hence we also conclude that the mass of the black hole is uncorrected, in agreement with the expectation of \cite{Smolic:2013gz}.

\subsection{Other terms in the FG expansion \label{fgterms}}

A feature of higher derivative gravity is the existence of additional degrees of freedom contained in the metric. This occurs because the equations of motion are no longer second order. These new degrees of freedom will appear as new terms in the asymptotic FG expansion, which according to  the usual AdS/CFT lore will represent new
operators in the dual CFT. Here we show how the entanglement first law can be used
to derive the FG expansion for these new modes, including a derivation of the conformal dimensions of the CFT operators
to which they couple.

Of course, the physical interpretation of these modes is unclear. First, they typically have negative norm indicating
that the boundary theory is no longer unitary \cite{holoc}, and second, their masses are typically at the string scale where the low energy effective field theory is unreliable.  Nonetheless, they do satisfy the equations of motion, so we can ask how they fit mathematically into our discussion of the first law.

These new modes appear as additional solutions to the first law constraint $\d S^{grav}_B = \d E^{grav}_B$. Previously, we argued that a metric perturbation of the form \eqref{metpert} that satisfies the first law relation  must have $\D =d$ and be related to the stress tensor expectation value as we described in the preceding section. Nevertheless, perturbations with $\D \neq d$, with $\Delta$ an arbitrary real number,  are also allowed, as long as they satisfy $\d S_B^{grav} =0$.

To show how this works explicitly, we consider the example of general $R^2$ gravity,
with Lagrangian given by eq.~\eqref{lagrrsq}. We consider a metric perturbation of the form \eqref{metpert}. The  $x$ integral in eq.~\eqref{Xvar2} is convergent as long as
 $\D > d -2$. Performing this integral, we find
\be
\d S^{grav} = \frac{\ell^{d-3} R^{\D}\Omega_{d-2}}{2 \Gn} \frac{\Gamma\left(\frac{d-1}{2}\right)\Gamma\left(\half (\D-d)+1\right)}{2 \, \Gamma\left(  \frac{\D+1}{2}\right)} \left( \hat h_{00}^{(\D)}\, a_T + h^{(\Delta)}\, a_S \right)
\ee
where we have defined
\be
h^{(\Delta)} \equiv h^{(\D) i}{}_i - h^{(\D)}_{00} \quad{\rm and}\quad \hat h_{\mu\nu}^{(\D)}\equiv
  h^{(\D)}_{\mu\nu} - \frac{1}{d}\, h^{(\D)}\ .
\ee
Further the constant factors are given by
\be
a_T =  \frac{\tl^2\D}{4 \ell^2 (1+\D )}  \left[2 d (a_2+a_3+d\,a_3 )+a_2( d-\D) \D +4 a_1 \left(3-d+d \D -\D ^2 \right)-\frac{\ell^2}{\tl^2}\right] \nonumber
\ee
\bea
a_S & =& \frac{\tl^2\D}{4 d \ell^2 (1+\D )}  \left[\vphantom{\frac{\ell^2}{\tl^2}}
2 (d-3) d (a_2+a_3+d\,a_3)- (a_2+d\,a_2 +4 d\,a_3)(d-\D) \D +\right. \nonumber \\ &&\qquad\qquad
\qquad\qquad \left.-4 a_1 \left(3-d+d \D -\D ^2\right)-(d-1)\frac{\ell^2}{\tl^2}\right] \ .
\eea
We can then satisfy the equation $\d S^{grav}_B = 0$ at leading order in $R$, the radius of the ball, by demanding that the constants $a_T,a_S$
vanish. This is the case for $\D =0$ and $\D = \D_{T,S}$,  where\footnote{Of course, only the $\D_{S,T}^+$ solutions are physical, since only for them does the $x$ integral converge. It is interesting though that the $\d S_B^{grav} =0$ constraint also knows about the non-normalizable modes in gravity, including the perturbation of the boundary metric, with $\Delta =0$.  }
\bea \label{dtds}
\D_T^\pm &=& \frac{d}{2}  \pm \sqrt{\frac{d^2}{4}+\frac{2 a_3 d (d+1)+ 2 d a_2 -4 a_1 (d-3)- \ell^2/\tl^2}{4 a_1+a_2}} \nonumber \\
\D_S^\pm &=& \frac{d}{2}  \pm \sqrt{\frac{d^2}{4} +\frac{  (d-1) \ell^2/\tl^2 - 2 (d-3) [2 a_1 + a_2 d + d(d+1) a_3]}{4 a_1+a_2 (d+1) +4 a_3 d}}
\eea
We have checked that these expressions agree with the coefficients of the asymptotic falloffs of solutions to the equations of motion in $R^2$ gravity.\footnote{For completeness, we reproduce  the equations of motion that follow from the Lagrangian \eqref{lagrrsq}:
\bea
\frac{\s}{\tl^2} G_{\mu\nu} -\frac{d(d-1)}{2\tl^4} g_{\mu\nu}  =  \half  \left[ a_1 R_{\mu\nu\rho\s} R^{\mu\nu\rho\s} + a_2 R_{\mu\nu} R^{\mu\nu} + a_3 R^2 - (a_2 + 4 a_3)\, \Box R \, \right] g_{\mu\nu} - 2 a_1 R_{\mu\a\b\g} R_\nu{}^{\a\b\g}  \nonumber \\
-   (2 a_2+4 a_1 ) R_{\mu\a\nu\b} R^{\a\b} - 2 a_3 R R_{\mu\nu}+ 4 a_1 R_{\mu\a} R_\nu{}^\a + (2 a_3 +a_2+2a_1) \nabla_\mu \nabla_\nu R - (a_2+4a_1) \Box R_{\mu\nu}  \nonumber
\eea
On the AdS solution of radius $\ell$, the relationship between $\ell$ and $\tl$ is given in eq.~\reef{scales}.} Also,
for $d=3$, $\D_S^+$ agrees with the operator dimension that was obtained in \cite{Smolic:2013gz}, also
 by solving the asymptotic equations of motion. Therefore, imposing $\delta S_{B}^{grav} = 0$ as $R \to 0$ ensures that the asymptotic equations of motion are satisfied, a claim which we use in section \ref{ss:bulkmod}.

\section{Discussion} \labell{discuss}

In this paper, we have seen that a universal relation between entanglement entropy and `modular' energy for small perturbations to the vacuum state of a CFT leads, in the holographic context, to a nonlocal constraint on the dual spacetimes, which is exactly equivalent to the linearized gravitational equations. Thus, given any holographic CFT, we can derive the linearized bulk equations knowing only the entanglement functional. Moreover, as we showed in sections \ref{ss:derivet} and \ref{hcurv}, we can also derive the asymptotic boundary conditions for the metric perturbation, as well as an expression for the holographic stress tensor. When matter couplings to curvature vanish, these results taken together imply that from the entanglement functional, we can derive the complete map from states to metrics at the linearized level
about the vacuum.

We have also shown that this non-local gravitational constraint is precisely the first law of black hole thermodynamics (in the form proved by Iyer and Wald) applied to certain Rindler patches of pure AdS that can be also interpreted as
 zero-mass hyperbolic black holes.
Thus, we have a result that
holds purely in
classical gravity: in any classical gravitational theory for which  anti-de Sitter space is a solution and for which the first law of black hole thermodynamics holds for some Wald functional $\Sw$, small perturbations about the AdS vacuum solution are governed by the linearized gravitational equations obtained from varying the Lagrangian associated to $\Sw$. This provides a converse to the theorem of Iyer and Wald, but also a microscopic understanding of the origin of the Iyer-Wald first law for AdS-Rindler horizons.

\subsubsection*{Relation to the work of Jacobson}

The results in this paper are reminiscent of (and partly motivated by) the work of Jacobson \cite{revelation} (see also \cite{ted2,ted3,Bianchi:2012br}). There, it was shown that if the first law of thermodynamics -- governing the local change in entropy (defined to be horizon area) as a certain bulk energy flows through the horizon --
 is assumed to hold for an arbitrary Rindler horizon,
 then the full nonlinear Einstein equations must be satisfied. In Jacobson's case, there was no microscopic understanding of the meaning of the entropy, and thus no fundamental understanding of why the thermodynamic relation should hold. By contrast, in our case there is a precise microscopic understanding of both the energy and the entropy appearing in our relation $\delta S_B = \delta E_B$, and a proof of the first law at the microscopic level. Also, our gravity analysis applies to an arbitrary higher curvature theory, a scenario that is problematic with Jacobson's approach \cite{ted3}. On the other hand, because our proof is based on global rather than local Rindler horizons, we were only able to obtain the gravitational equations of motion at the linearized level.

\subsubsection*{Deriving the nonlinear equations?}

It is obviously interesting to ask whether we can extend our results to the nonlinear level. On the CFT side, the entanglement entropies for finite perturbations to the vacuum state are still constrained by the modular energies, but the constraint is the {\it inequality} $\Delta S_A \le \Delta \langle H_A\rangle$ following from the positivity of relative entropy. For any ball-shaped region, we can still translate this inequality to a constraint on the bulk metric. The set of all such constraints should significantly restrict the allowed bulk spacetimes, but it seems unlikely that these restrictions will fully determine the bulk equations at the nonlinear level. In particular, the nonlinear gravitational equations are sensitive to all the other fields present in the classical bulk theory, including the components of the metric along any extra compact directions. These additional degrees of freedom depend significantly on which holographic CFT we are considering. Thus, starting from the universal relation $\Delta S_A \le \Delta \langle H_A\rangle$ (or any other universal relation for holographic CFTs) one might realistically expect to recover only a part of the constraints implied by the full non-linear equations; for example, one might obtain
Einstein's equations with the additional assumption that no other matter fields are turned on in the bulk.

Another interesting possibility is that one might be able to obtain some constraints at the nonlinear level in the bulk even from the linearized entanglement first law, by considering bulk perturbations which are kept finite but taken to be localized closer and closer to the
AdS-Rindler horizon.
In such a limit, the energy perturbation in the CFT vanishes due to gravitational redshift effects. By considering infinitesimal perturbations away from this limit, the linearized CFT first law should apply, but on the gravity side, it would appear that we will obtain constraints on a finite perturbation localized near the horizon. This may be closely related to the approach of Jacobson.

\subsubsection*{Quantum first law in the bulk}

Finally, it would be interesting to understand the implications of the entanglement first law (in its infinitesimal form) beyond the classical level on the gravity side. Since the entanglement first law is an exact relation, it can also be used to study subleading quantum gravitational corrections to the classical results that we have derived, or CFT states that do not have a classical bulk interpretation. These quantum states/corrections can be easily identified by the scaling of their energy and entropy  with the
  central charge in the CFT: while the classical contributions are proportional to the central charge, the quantum ones scale with a lower power  of it. Thus, the first law should place constraints on the quantum behaviour of the bulk gravitational theory
 and
   will likely also  involve an understanding of the quantum corrections to the Ryu-Takayanagi formula as discussed recently in \cite{Barrella:2013wja,Faulkner:2013ana,damian}.

\section*{Acknowledgements}

We thank Horacio Casini, Nima Lashkari, Aitor Lewkowycz, Juan Maldacena, Don Marolf, and Sasha Zhiboedov,  for useful conversations.
TH, RCM, and MVR also acknowledge the support of the KITP during the program
``Black Holes: Complementarity, Fuzz, or Fire?" where some of this work was
done. The research of TH is supported in part by the National Science
Foundation under Grant No. NSF PHY11-25915. Research at Perimeter Institute is
supported by the Government of Canada through Industry Canada and by the
Province of Ontario through the Ministry of Research \& Innovation. The research of MVR and RCM is supported in part by the Natural Sciences and Engineering Research Council of Canada. RCM also
acknowledges support from the Canadian Institute for Advanced Research.  The research of MG is  supported by
the DOE grant DE-SC0007901. TF is supported by NSF Grant No. PHY-1314311.

\appendix

\section{Vanishing of the integrand}\labell{app:slide}

Suppose
\be\labell{rds}
\int_{\Sigma}\!\! d^{d-1}x\,dz\ f(\vec{x},z) = 0 \quad\;\;\; \forall R,\vec{x}_0
\ee
where $\Sigma(R,\vec{x}_0)$ is the region $z\geq 0$,  $|\vec{x} - \vec{x}_0|^2 + z^2 \leq R^2$. We would like to show that \eqref{rds} implies that $f = 0$. To prove this, differentiate the integral, and define
\be
I_R = \p_R \int_{\Sigma}\!\! d^{d-1}x\,dz\ f =0\ \; , \;\;\;\; \quad I_i = \p_{x_0^i} \int_{\Sigma}\!\! d^{d-1}x\,dz\ f=0 \ .
\ee
These are the average and the first moment of $f$ on the hemisphere $\tilde{B}(R,x_0)$,
\bea
I_R = \int_{\tilde{B}}  \!\!  dA \ f= 0 \;, \;\;\;\; \quad
I_i = \int_{\tilde{B}} \!\!  dA \ x^i\, f = 0 \label{radon}
\eea
where $dA$ represents the area element on $\tilde B$.
Now we can repeat the argument replacing $f \to x^i f$ in (\ref{rds}), and deduce that  all moments of $f$ vanish on every hemisphere $\tilde{B}$.  We conclude that $f=0$, as we needed to show.

An alternative argument for the vanishing of $f$ is to note that the integral in (\ref{radon}), viewed as a map from $\tilde{B}$ to $\mathbb{R}$, defines the ``hyperbolic Radon transform'' of the function $f$, whose vanishing implies the vanishing of the function, assuming that $f$ is continuous \cite{radon}.

\section{Noether identities and the off-shell Hamiltonian}\labell{s:noether}

In this section, we derive the Noether identities for diffeomorphism invariance, and show that $\bJ[\xi] = d\bQ[\xi] + \xi^a \bC_a$ as claimed in (\ref{jcon}).

Under a diffeomorphism, the variation of the action $I$ is
\be
\delta_\xi I = \int \bepsilon (  E^{\phi}\delta_\xi \phi )
\ee
with the sum over fields $\phi$ implicit. The integrand for a field of rank $r$ is
\bea 
\bepsilon\, (E^\phi)_{a_1\cdots a_r}^{b_1\cdots b_s}\, \delta_\xi \phi^{a_1\cdots a_r}_{b_1\cdots b_s}
 &=& \bepsilon\,  (E^\phi)_{a_1\cdots a_r}^{b_1\cdots b_s}\,\left(\xi^b  \del_b \phi^{a_1\cdots a_r}_{b_1\cdots b_s} - \sum_{i=1}^r \del_\lambda \xi^{a_i} \phi^{a_1 \cdots \lambda \cdots a_r}_{b_1\cdots b_s}
+\sum_{i=1}^s \del_{b_i} \xi^{\lambda} \phi^{a_1 \cdots  a_r}_{b_1\cdots\lambda\cdots b_s}
\right)\nonumber\\
&=& \bepsilon \xi^b (E^\phi)_{a_1\cdots a_r}^{b_1\cdots b_s} \del_b \phi^{a_1\cdots a_r}_{b_1\cdots b_s} 
+ \bepsilon \xi^b \sum_{i=1}^r \del_\lambda\left[ (E^\phi)_{a_1 \cdots  b \cdots a_r}^{b_1\cdots b_s}
 \phi^{a_1 \cdots \lambda \cdots a_r}_{b_1\cdots b_s}\right] 
\nonumber\\
&&\qquad-\bepsilon \xi^b \sum_{i=1}^s \del_{b_i}\left[ (E^\phi)_{a_1 \cdots  a_r}^{b_1\cdots b_s}
 \phi^{a_1 \cdots  a_r}_{b_1\cdots b\cdots b_s}\right]  - d(\xi^a \bC_a)\label{pdeom}
\eea
where the dots indicate that indices appear in the $i$th position, and the constraints $\bC_a$ are defined in eq.~(\ref{defc}). If $\xi$ has compact support, then the total derivative does not contribute and since $\delta_\xi I = 0$ for any $\xi$, we have the following identity for the integrand,
\be\labell{dele}
\sum_\phi \left( (E^\phi)_{a_1\cdots a_r}^{b_1\cdots b_s} \del_b \phi^{a_1\cdots a_r}_{b_1\cdots b_s} 
+ \sum_{i=1}^r \del_\lambda\left[ (E^\phi)_{a_1 \cdots  b \cdots a_r}^{b_1\cdots b_s}
 \phi^{a_1 \cdots \lambda \cdots a_r}_{b_1\cdots b_s}\right] 
- \sum_{i=1}^s \del_{b_i}\left[ (E^\phi)_{a_1 \cdots  a_r}^{b_1\cdots b_s}
 \phi^{a_1 \cdots  a_r}_{b_1\cdots b\cdots b_s}\right] \right) = 0 \ .
\ee
This is the Noether identity.

Next, remember that the Noether current \eqref{defj} satisfies $d\bJ[\xi] = -\bepsilon E^\phi \delta_\xi \phi $. Using \eqref{pdeom} and  the Noether identity, this becomes
\be
d\bJ[\xi] = d(\xi^a \bC_a)
\ee
for all diffeomorphisms $\xi$. It follows that \cite{waldlemma}
\be
\bJ [\xi] = d\bQ[\xi] + \xi^a \bC_a  \ ,
\ee
for some $\bQ$, which we take to be the off-shell definition of the Noether charge $\bQ$.

\section{Example: Einstein Gravity coupled to a Scalar}\labell{app:einsteinscalar}
In this appendix we review the covariant formalism applied to Einstein gravity coupled to a 
scalar field. The Lagrangian is
\be
\bL = \bepsilon\left[ \frac{1}{16\pi \Gn}R - \frac{1}{2} (\p \psi)^2 - V(\psi)\right] \ .
\ee
The cosmological constant is included in the scalar potential $V(\psi)$. The definitions (\ref{varL}) and (\ref{defj}) give
\be\labell{thetae}
\bTheta = \left[\frac{1}{16\pi \Gn}\left( \del_b \delta g^{ab} - \del^a \delta g_b^{\ b}\right) - \delta \psi \del^\a \psi\right]\bV_a
\ee
and
\be
\bJ = \left[\frac{1}{8\pi \Gn}\del_e\left( \del^{[e}\xi^{d]}\right) +2 (E^g)^d_{\ e}\xi^e\right]\bV_d
\ee
where $E^g$ is the gravitational equation of motion,
\be
E^g_{ab} = \frac{1}{16 \pi \Gn}\left(R_{ab} - \frac{1}{2}g_{ab} R\right) -\half \p_a \psi \p_b \psi +\half g_{ab}\left[\frac{1}{2}(\p\psi)^2 - V(\psi)\right] \ .
\ee
The Noether current can be written
\be\labell{einj}
\bJ = d\bQ  + 2\xi^a E^g_{ab} \bV^b
\ee
where
\be
\bQ = -\frac{1}{16 \pi}\del^a \xi^b \bV_{ab} \ .
\ee

\section{Form of the bulk charge \labell{app:formcharge}}

In this appendix, we show that the linearized modular energy defined by the bulk Wald-Noether
procedure always take the simple form noted in eq.~\eqref{de1}. We start
with eq.~\eqref{ewald}, reproduced here for convenience:
 \be
\delta E^{grav}_{(1)} = \int_{B} \left( \delta \bQ[\xi_B] - \xi_B \cdot \bTheta(\delta \phi) \right) \label{ewaldprime}
 \ee
where the Killing vector $\xi_B$ is given in eq.~\eqref{defxi}. Into this equation
we would like to substitute the asymptotic form of the metric perturbation \eqref{pge}, representing the stress tensor perturbation.
As we argued in the main text, modes with different falloffs  will not contribute, since
they have the wrong scaling dimension.

As shown in \cite{iyer1}, the most general form of $\bQ[\xi]$  is
\be \label{qiw}
\bQ[\xi] = \mathbf{X}^{cd} \nabla_{[c} \xi_{d]} +  \mathbf{W}_c \xi^c + \mathbf{Y}(\phi,\mathcal{L}_\xi \phi) + d\mathbf{Z}(\xi,\phi) 
\ee
where $\mathbf{Y}$ is linear in $\mathcal{L}_\xi \phi$, $\mathbf{Z}$ is linear in $\xi$, and all forms are covariant expressions constructed from the fields. We assume there is no matter with linear couplings to curvature. The general covariant form of $\mathbf{X}^{cd}$ is
\be
\mathbf{X}^{cd} = X^{abcd} \bepsilon_{ab}
\ee
where $X^{abcd}$ is antisymmetric in both is first two and last two indices. Using symmetry and arguments similar to those in section \ref{hcurv}, at zeroth and first order around AdS and to leading in the $z$ expansion, we must have (ignoring coefficients)
\be
\left. X^{abcd}\right|_{AdS}  \propto g^{\langle ab} g^{cd\rangle} \;, \;\;\;\;\; \d X^{abcd}  \propto  g^{\langle ab} g^{cd\rangle} h + h^{\langle ab} g^{cd\rangle} \ . \label{linx}
\ee
The contribution of the first term in eq.~\eqref{qiw} to $\delta E_{(1)}^{grav}$ is then
\be
I_X = \int_B \d \left(X^{abcd} \bepsilon_{ab} \nabla_{c} \xi_{d} \right) = \int_B \left(\d X^{abcd}\,  \nabla_{c} \xi_{d} + \half\,  h\,   X^{abcd} \,  \nabla_{c} \xi_{d} +  X^{abcd}  \, \d (\nabla_{c} \xi_{d}) \right) \bepsilon_{ab}
\ee
where the quantities without $\d$'s are evaluated on the background AdS solution. The non-zero background components are
\be
\e_{ab} \r \e_{tz} \propto  \frac{d^{d-1} x}{z^{d+1}} \;, \;\;\;\;\; \nabla_{[i} \xi_{t]} = \frac{x^i}{R z^2} \;, \;\;\;\;\; \nabla_{[z} \xi_{t]} = \frac{R^2-|\vec{x}|^2}{2 R z^3} \;, \;\;\;\;\;  X^{tz cd} \propto z^4 \d^{[c}_t \d^{d]}_z \ .
\ee
Using eq.~\eqref{linx}, the leading behaviour of the linearized quantities reads
\be
 \d X^{tztz}  \propto z^6 (h_{tt}+ h_{\mu\nu}\eta^{\mu\nu}) \;, \;\;\;\;\; \d X^{tzti} \propto z^7 (\p_i h_{tt} + \ldots) \;, \;\;\;\;\; \d ( \nabla_{[z} \xi_{t]}) \propto \zeta^t \p_z h_{tt}
\ee
It is clear from the above expressions that only the leading  terms in  $X^{tztz}$ and $\d X^{tztz}$  will contribute as $z \r 0$. Plugging in the $z$-dependence of $h_{\mu\nu}$, one finds that all  the non-vanishing contributions are proportional to $h^{(d)}_{tt}$ or $h^{(d)}_i{}^i$. Requiring moreover that $h^{(d)}_\mu{}^\mu =h^{(d)}_i{}^i - h_{tt}^{(d)} = 0$, which follows from tracelessness of the CFT stress tensor, we find that
\be
I_X  \propto \int d^{d-1} x \, h_{tt}^{(d)} \zeta^t
\ee
where we used the fact that $\lim_{z \r 0} \xi^t = \zeta^t$.

The contribution of the $\mathbf{W}_c \xi^c = \mathbf{W}_t \zeta^t $ term is easy to evaluate, taking into account the fact that the $d-1$ form $\mathbf{W}_c$ is a covariant expression constructed from $h_{ab}$, $g_{ab}$ and their background covariant derivatives. The most general form of  $\mathbf{W}_c$, linearized around AdS, is thus
\be
\mathbf{W}_c = \bepsilon_{ab} \mathcal{F}^{ab}{}_c  
\;, \;\;\;\;\; \mathcal{F}^{ab}{}_c = f_1(\Box) (\nabla^a h^{b}{}_c - \nabla^b h^{a}{}_c ) + f_2(\Box) (\d^a{}_c \nabla^b h - \d^b{}_c \nabla^a h) \ .
\ee
The only non-zero contribution on $B$ will be from $\mathcal{F}^{[tz]}{}_t$, and using tracelessness of the leading term in $h$ one can easily show that
\be
\int_B \mathbf{W}_c \, \xi^c \propto \int_B d^{d-1} x \, h_{tt}^{(d)} \zeta^t \ .
\ee
The $\xi \cdot \bTheta$ term in eq.~\eqref{qiw} has the same form as $\bW_c\xi^c$ so can be treated similarly.

The term $\mathbf{Y}$ in eq.~\eqref{qiw} comes from the ambiguity $\bTheta \to \bTheta + d\mathbf{Y}(\delta\phi)$.  Together, these terms contribute to $\bchi$ in the combination
\be
\delta \bY(\delta_\xi \phi) - \delta_\xi \bY(\delta \phi) \ .
\ee
This vanishes for a background Killing vector.
%

Finally, the $d \mathbf{Z}$ term is an ambiguity that comes from the fact that $\bQ$ is only defined by its derivative.  We fix this ambiguity to zero by requiring that there are no boundary terms in the horizon entropy.

 The overall conclusion is
 \be
 \delta E^{grav}_{(1)}  = C_1 \int_\B d^{d-1} x \, h_{tt}^{(d)} \zeta^t \ ,
 \ee
 for some constant $C_1$.

\end{document}